\documentclass[aps,prl,reprint,amsmath,superscriptaddress]{revtex4-1}
\usepackage{amsmath}
\usepackage{txfonts}
\usepackage{hyperref}
\usepackage{graphicx}
\usepackage{float}
\usepackage{xcolor}
\usepackage[normalem]{ulem}
\usepackage{todonotes}
\usepackage[utf8]{inputenc}
\usepackage{bm}
\usepackage{siunitx}
\usepackage{pdfpages}
\usepackage{etoolbox} 

\makeatletter
\patchcmd{\@outputpage@head}{\@ifx{\LS@rot\@undefined}{}{\LS@rot}}{}{}{}
\makeatother

\DeclareMathOperator{\Prob}{Prob}

\newcommand{\Dz}{D z}

\begin{document}
\title{Edge of chaos and avalanches in neural networks with heavy-tailed synaptic weight distribution}

\author{\L{}ukasz Ku\'smierz}
\email{nalewkoz@gmail.com}
\affiliation{Laboratory for Neural Computation and Adaptation, RIKEN Center for Brain Science, 2-1 Hirosawa, Wako, Saitama 351-0198, Japan}
\author{Shun Ogawa}
\affiliation{Laboratory for Neural Computation and Adaptation, RIKEN Center for Brain Science, 2-1 Hirosawa, Wako, Saitama 351-0198, Japan}
\author{Taro Toyoizumi}
\affiliation{Laboratory for Neural Computation and Adaptation, RIKEN Center for Brain Science, 2-1 Hirosawa, Wako, Saitama 351-0198, Japan}
\affiliation{Department of Mathematical Informatics, Graduate School of Information Science and Technology, The University of Tokyo, Tokyo 113-8656, Japan}

\begin{abstract}
    We propose an analytically tractable neural connectivity 
    model with power-law distributed synaptic strengths. 
    When threshold neurons with biologically plausible number of incoming 
    connections are considered, our model features a continuous transition 
    to chaos and can reproduce biologically relevant low activity levels 
    and scale-free avalanches, 
	i.e. bursts of activity with power-law distributions of sizes and lifetimes. In contrast, the Gaussian counterpart exhibits a discontinuous 
	transition to chaos and thus cannot be poised near the edge of chaos. We validate our predictions in simulations of networks of binary as 
    well as leaky integrate-and-fire neurons. 
    Our results suggest that heavy-tailed synaptic distribution 
    may form a weakly informative sparse-connectivity 
    prior that can be useful in biological and artificial adaptive systems.
\end{abstract}

\maketitle
Scale-free neuronal avalanches, commonly associated with criticality, 
have been observed in cortical networks in various settings, including 
cultured and acute slices from rat somatosensory cortex
\cite{beggs2003neuronal}, 
eye-attached \textit{ex vivo} preparation of turtle visual cortex 
\cite{shew2015adaptation}, 
visual cortex in anesthetized rats 
\cite{fontenele2019criticality},
primary visual cortex in anesthetized monkeys
\cite{fontenele2019criticality}, 
and premotor, motor, and somatosensory cortex in awake monkeys 
\cite{petermann2009spontaneous}. 
Criticality implies the existence of a continuous transition between two 
distinct collective phases. 
In the context of neuronal avalanches, 
most commonly studied transitions are between quiescent and active states 
\cite{beggs2003neuronal,haldeman2005critical,kinouchi2006optimal} 
or synchronuous and asynchronuous states 
\cite{di2018landau, fontenele2019criticality}. 
In addition to providing a plausible generating mechanism for the 
neuronal avalanches, the existence of a continuous transition would have 
important functional implications, as it has been shown that 
computation is most efficient around a critical point \cite{kinouchi2006optimal,stoop2016auditory,munoz2018colloquium}, 
often associated with the \textit{edge of chaos} 
\cite{langton1990computation,bertschinger2004real,legenstein2007makes,toyoizumi2011beyond,schuecker2018optimal}. 
However, the relation between neuronal avalanches, criticality, and edge of chaos 
is not fully understood \cite{benayoun2010avalanches,beggs2012being,munoz2018colloquium}. 
	
Different scenarios of the transition to chaos in randomly connected neural networks 
were extensively studied over the last $30$ years 
\cite{sompolinsky1988chaos,molgedey1992suppressing,amit1997model,brunel2000dynamics,rajan2010stimulus,massar2013mean,stern2014dynamics,aljadeff2015transition,kadmon2015transition,wainrib2016local,crisanti2018path,schuecker2018optimal}. 
According to the prevailing assumption rooted in the central limit theorem, 
the total synaptic input current of each neuron can be modeled 
as a Gaussian random variable (\textit{Gaussian assumption}).  
Here we argue that the Gaussian assumption cannot account for 
some of the experimentally observed features of neuronal circuits. 

In particular, the continuous nature of the phase transition observed 
in the conventional models  
is sensitive to theoretical assumptions that are not biologically grounded. 
Most works that study transition to chaos employ rate models with 
continuous non-thresholded activation functions, often of a sigmoidal shape 
\cite{sompolinsky1988chaos,molgedey1992suppressing,rajan2010stimulus,stern2014dynamics,aljadeff2015transition,crisanti2018path,schuecker2018optimal}. 
Sometimes thresholds are introduced, but often the analysis is  
restricted to the suprathreshold regime 
\cite{kadmon2015transition, harish2015asynchronous}.
But most neurons in the brain spike only when driven by strong enough 
excitatory synaptic input above a threshold \cite{dayan2001theoretical,hausser2003less,gerstner2014neuronal}. 
Thus, we model a self-sustained (autonomous) activity 
in a network of individually subthreshold neurons. 
Other models exhibiting continuous transition to chaos 
\cite{bertschinger2004real} 
or neuronal avalanches 
\cite{kinouchi2006optimal,millman2010self} 
rely on extremely sparse (${\sim}10$ connections per neuron) networks.
However, many neurons in the vertebrate brain receive a 
large number of inputs from 
other cells (${\sim}10^4$) \cite{dayan2005book}. 
We observed that the transition to chaos becomes discontinuous when 
densely connected subthreshold units are used in tandem with 
the Gaussian assumption (Fig.~\ref{fig:spiking}) 
\cite{brunel2000dynamics, marshel2019cortical}. 
This discontinuous character of the transition
makes it hard for the network to robustly exhibit the edge of chaos, 
low activity levels, or avalanches. 
Although a discontinuous transition can be smoothed by noise, leading 
again to critical behavior away from the edge of chaos if the noise level is appropriate
\cite{scarpetta2018hysteresis}, 
such noise-induced criticality requires extra fine-tuning. 
We explore instead the possibility that an autonomous network exhibits critical behavior 
at the edge of chaos. 

%%%% FIGURE 1 %%%%%
\begin{figure}[htb!]
    \centering
    \includegraphics[trim={1.0cm 1.0cm 1.0cm 1.0cm},clip, width=1\linewidth]{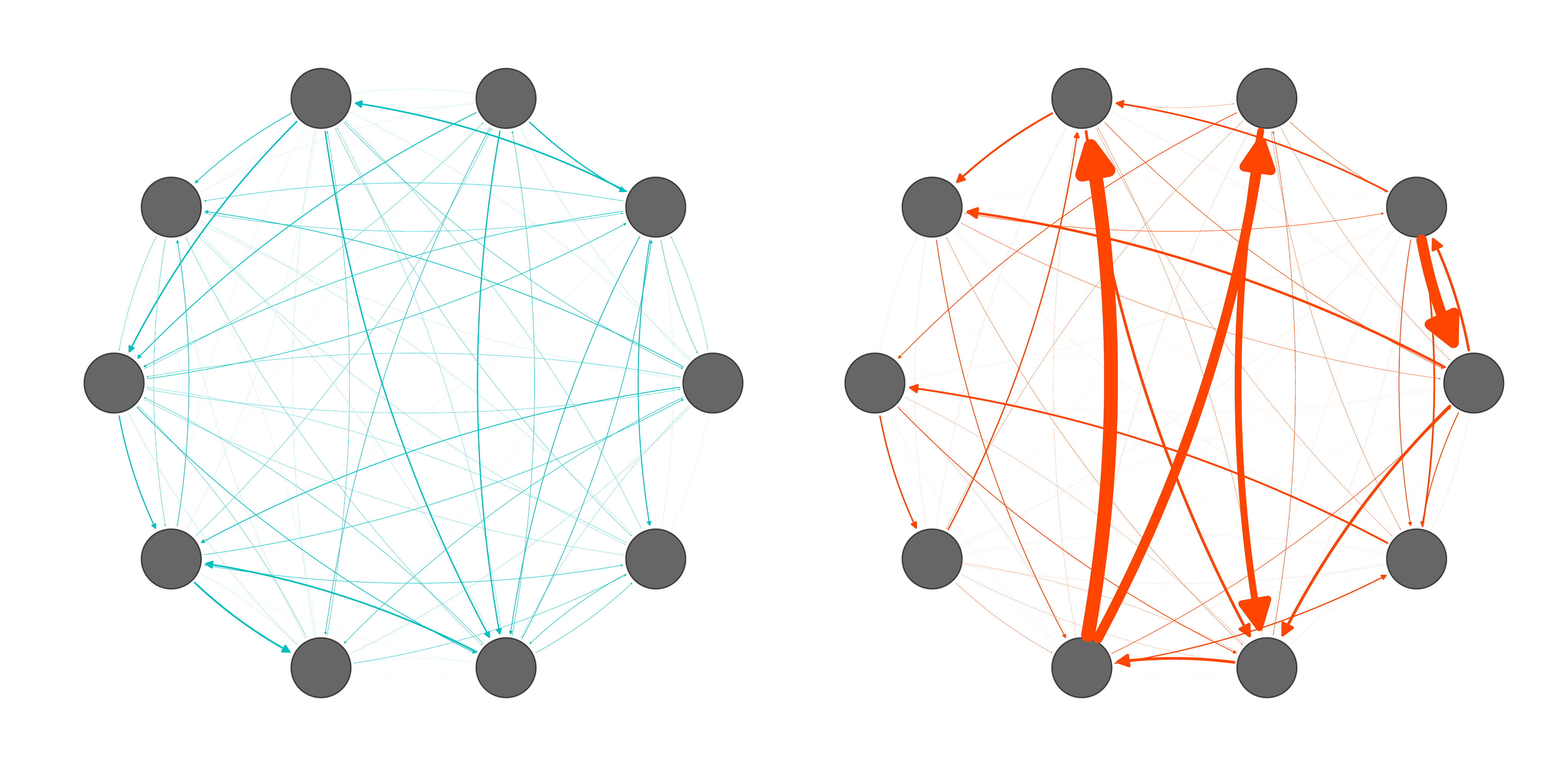}
    
    \includegraphics[width=0.9\linewidth]{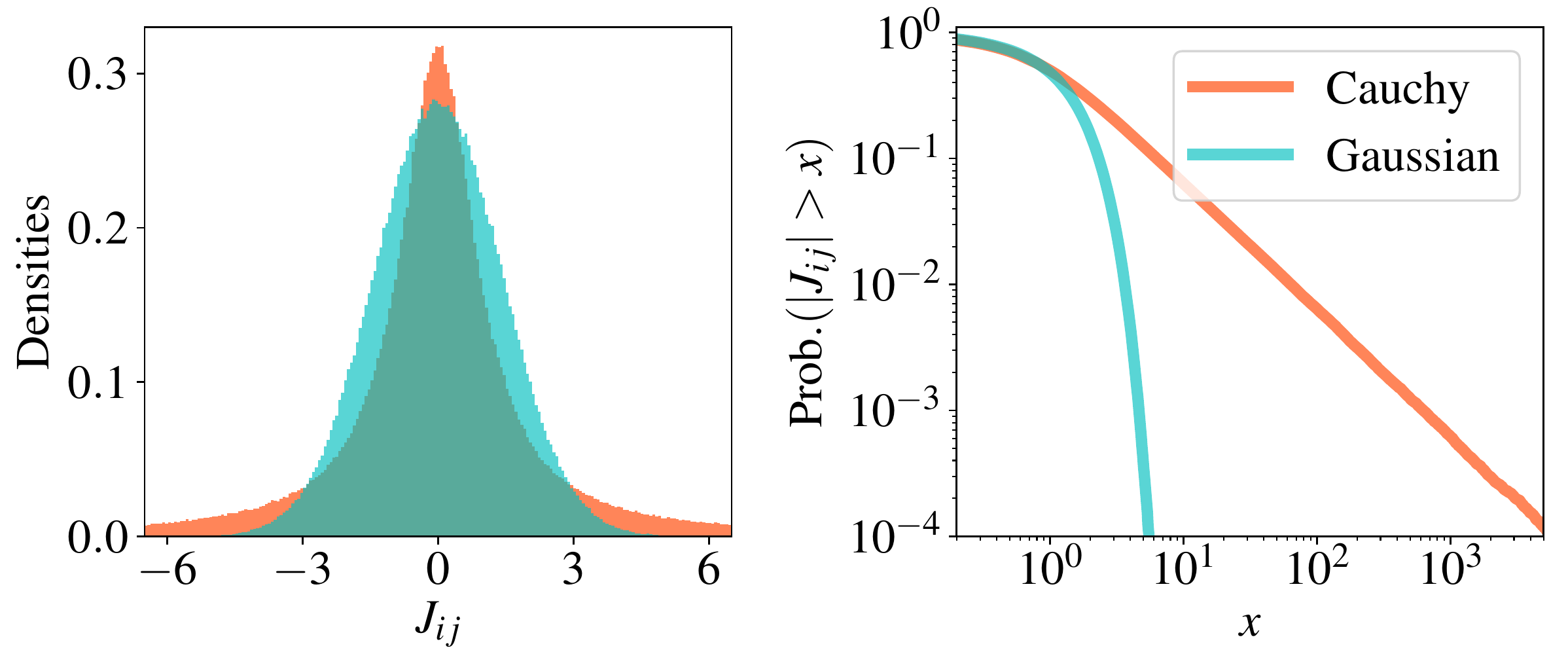}
    \includegraphics[trim={6.0cm 1.0cm 4.5cm 0.1cm},clip,angle=90,origin=c,width=0.46\linewidth]{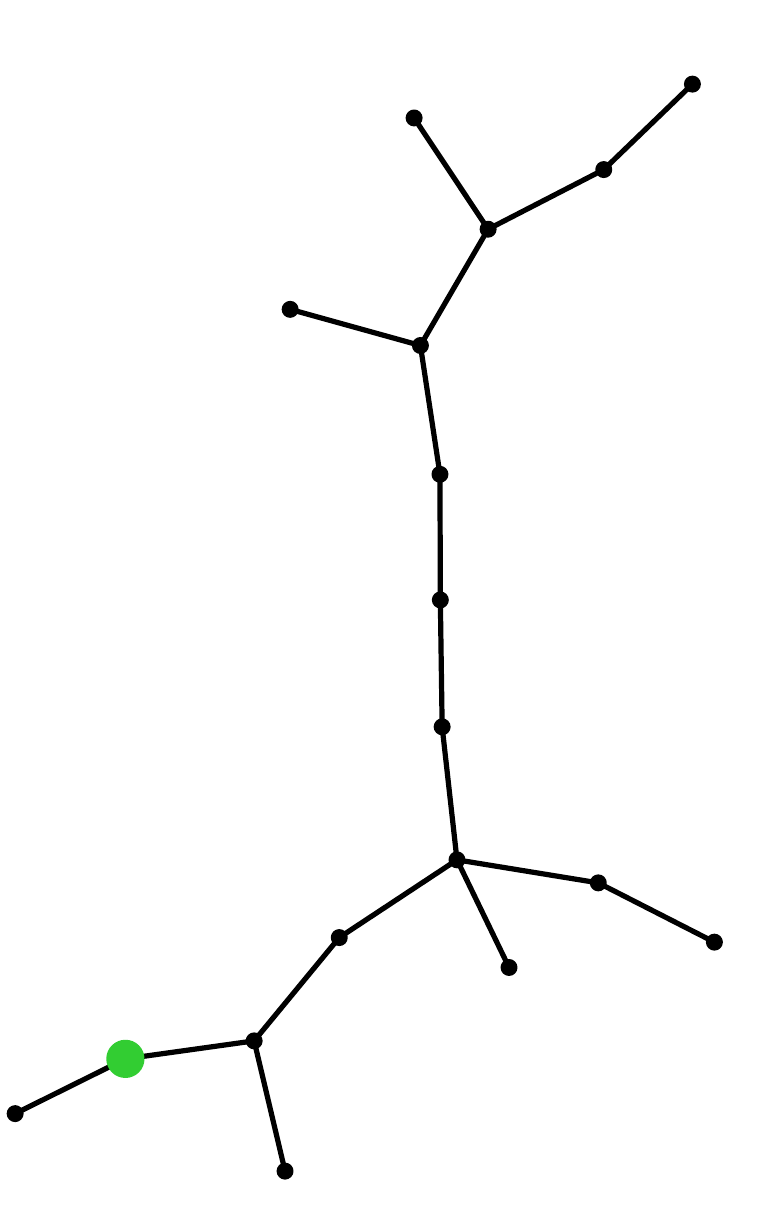}
    \includegraphics[trim={1.8cm 1.0cm 0.5cm 0.5cm},clip,width=0.53\linewidth]{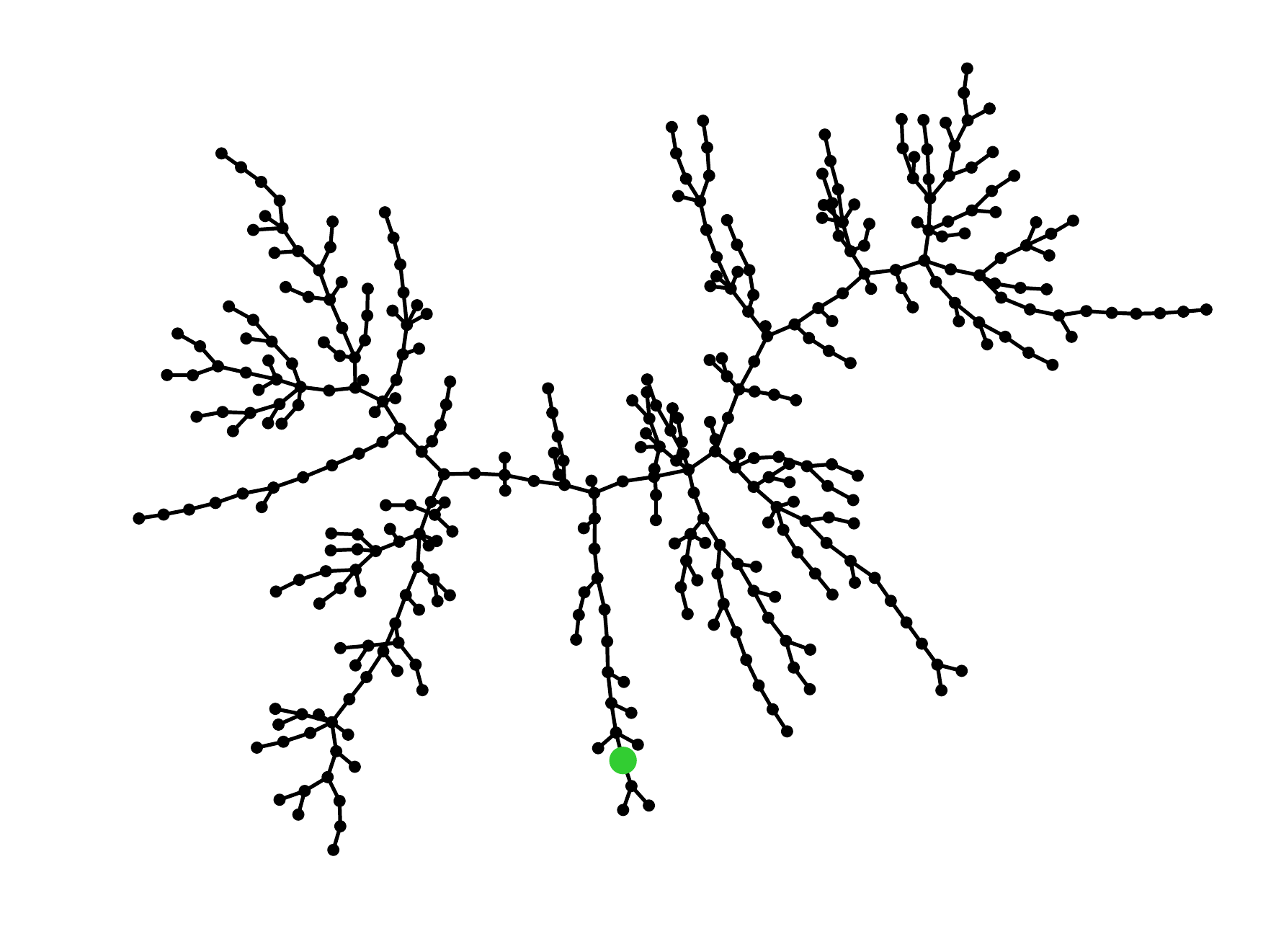}
    \caption{(top): Visualizations of neural networks with Gaussian (cyan) 
    and Cauchy (orange) distribution of weights. 
    Thickness and color saturation of edges correspond to the 
    (nonlinearly transformed) strengths of the connections. 
    (middle): Probability density functions (left) and 
    cumulative distribution functions (right) of Cauchy 
    and Gaussian random variables. 
    The Cauchy distribution features much thicker tails 
    than the Gaussian distribution. 
    (bottom): Sample realizations of the Poisson critical 
    branching process with the duration 
    $T=11$, size $S=18$ (left), and $T=35$, $S=367$ (right).
    The initial seeds are marked with the green color.
    In this work we show that activity of a fully connected Cauchy
    (but not Gaussian) network around the critical point 
    can be mapped to the critical branching process. 
    }
    \label{fig:nn}
\end{figure}

To fix this issue, we draw on the experimental works reporting heavy-tailed 
distributions of synaptic weights in various areas of the brain
\cite{sayer1990time,feldmeyer1999reliable,song2005highly,lefort2009excitatory,ikegaya2012interpyramid,loewenstein2011multiplicative}. 
Multiple theoretical mechanisms have been suggested to realize such distributions, 
e.g. 
modified spike-timing-dependent plasticity (STDP) rule 
\cite{gilson2011stability} or
STDP combined with homeostatic plasticity 
\cite{zheng2013network}. 
Notably, recent studies have suggested that experimentally observed activity-independent intrinsic spine dynamics can straightforwardly explain 
the heavy-tailed distributions of synaptic weights
\cite{yasumatsu2008principles, nagaoka2016abnormal,humble2019intrinsic,ishii2018vivo, okazaki2018calcineurin}. 

Although extensively studied, the computational role of synaptic heavy tails 
is still not fully understood. 
A log-normal distribution is often assumed and the results 
are obtained by means of computer simulations 
\cite{teramae2012optimal,ikegaya2012interpyramid, buzsaki2014log,teramae2014computational,omura2015lognormal}. 
Unfortunately, the log-normal distribution is not a stable distribution 
\cite{feller1971introduction}. 
Consequently, the corresponding distribution of the membrane potential 
depends in a nontrivial way on the details of the connectivity, 
including number of incoming connections. 
Moreover, if the number of incoming connections is scaled linearly with the number of neurons, 
the Gaussian assumption is recovered in the thermodynamic limit. 
This hinders theoretical approaches to study the effects of heavy-tails in these models. 
Therefore, a simple model that robustly predicts the effects of synaptic heavy tails is needed. 
We fill this gap by assuming random, power-law distributed synaptic weights.

Our aim is to inspect how the distribution of synaptic efficacies, 
modulated by the activation function, affects 
the transition to chaos and the associated avalanches. 
To this end, in our calculations we focus on the network effects and hence 
simplify the dynamics of individual neurons by considering 
the following discrete-time network dynamics
    \begin{equation}
         x_i(t+1) = \sum\limits_{j=1}^N J_{ij} \phi(x_j(t))
         ,
         \label{eq:general-network-dynamics}
    \end{equation}
where $\phi(x)$ is the activation function, 
assumed to be identical across the network, 
and $\bm{J}$ is the connectivity matrix. 
The network is fully connected and the synaptic weights 
are independently drawn from  the common Cauchy distribution (Fig.~\ref{fig:nn})
\begin{equation}
    \rho(J_{ij}) = \frac{1}{\pi}\frac{g/N}{(g/N)^2 + J_{ij}^2}
    \label{eq:cauchy-dist},
\end{equation} 
with the characteristic function
\begin{equation}
\Phi_{J}(k) = 
e^{-\gamma |k|},
\label{eq:general-cauchy-cf}
\end{equation}
where $\gamma = g/N$
defines the width of the distribution. 
We refer to the model prescribed by 
(\ref{eq:general-network-dynamics}) and (\ref{eq:cauchy-dist}) 
as the \textit{Cauchy network}. 

Due to the generalized central limit theorem 
\cite{feller1971introduction}, 
in the thermodynamic limit of $N\to\infty$ 
results obtained for the Cauchy model are applicable to networks 
with connections drawn independently from any symmetric  
distribution with $1/x^2$ tails 
that are scaled with the number of neurons as $1/N$. 
In contrast, in the more commonly used \textit{Gaussian networks},
the synaptic weights are independently drawn from the normal 
distribution $J_{ij} \sim \mathcal{N}(0, g^2/N)$. 
In the thermodynamic limit this corresponds to connectivity matrices 
with entries independently drawn from any distribution with zero mean and 
a finite variance, as long as the weights are scaled as $1/\sqrt{N}$.

The natural order parameter in the system at hand is 
the mean network activity, defined as 
\begin{equation}
    m_t = 
    \frac{1}{N}\sum\limits_{i=1}^N
        | \phi(x_i(t))|.
    \label{eq:def-order-parameter}
\end{equation} 
The state at time $t+1$ depends on $\bm{J}$ and $\bm{x}(t)$. 
We fix the activity vector at time $t$ and treat $\bm{x}(t+1)$ as 
a function of $\bm{J}$, 
which allows us to characterize the distribution of $x_i(t+1)$ 
using $\Phi_{x_i(t+1)}(k)$ as
\begin{equation}
    \left\langle
    e^{
    i k x_i(t+1)
    }
    \right\rangle_{\bm{J}}
    =
    \exp\left(
    -g |k| N^{-1}\sum_{j=1}^N 
    |\phi(x_j(t))| 
    \right)
    =
    \exp\left(
    -g m_t |k|
    \right)
    .
    \label{eq:cf_x}
\end{equation}
The activity of a neuron at time $t+1$, as a function of synaptic weights, 
is a Cauchy random variable whose width depends on the activity at time $t$ 
only through its mean value. 

To proceed we assume self-averaging, i.e. that the mean activity 
is the same for each realization of the network. 
Since in our model synaptic weights are statistically the 
same for all neurons, 
in the limit of ${N\to\infty}$ the mean activity can alternatively be expressed as
\begin{equation}
    m_t = \langle |\phi(x_i(t))| \rangle_{\bm{J}} \;\; {(\forall i)}.
    \label{eq:m_alternative}
\end{equation} 
We use the result of (\ref{eq:cf_x}), i.e. 
that $x_i(t)$ averaged over $\bm{J}$ is a Cauchy variable with 
$\gamma = gm_t$, together with (\ref{eq:m_alternative}) 
and arrive at the evolution of the mean activity 
in a simple integral form 
\begin{equation}
    m_{t+1} = 
    \int\limits_{-\infty}^{\infty} \Dz 
    |\phi(g m_t z)|,
    \label{eq:general-sc-mf}
\end{equation}
where 
${\Dz=\pi^{-1} \mathrm{d}z/(1 + z^2)}$
denotes that the integral is calculated with respect to the standard Cauchy measure. 
The steady-state mean activity can be obtained from (\ref{eq:general-sc-mf}) 
in a self-consistent manner. 
% =====================================================================================================

We are now in the position to analyze the dependence of the dynamics of the Cauchy 
network on the activation function. 
For $\phi(x)=x$, the integral on the right-hand side (RHS) of (\ref{eq:general-sc-mf}) 
diverges, suggesting that the network is unstable. 
Indeed, it is easy to understand why this is the case. 
For linear networks the dynamics is fully determined by
the eigenvalues of the connectivity matrix $\bm{J}$. 
It is known that, in contrast to random matrices with Gaussian entries,
a Cauchy random matrix 
features an unbounded support of the eigenvalues density, 
even in the limit of $N\to\infty$ 
\cite{cizeau1994theory,burda2002free,gudowska2020synaptic}. 
Thus, we can conclude that, regardless of the value of the $g$,
the dynamics of a Cauchy neural network is in this case divergent.
For the same reason, any $\phi(x)$ that is linear around 
$x\approx 0$ and grows sufficiently slow for large $x$ 
leads to a self-sustained, active 
dynamics for any $g$ \cite{SI}. 
However, in the biologically relevant regime neurons exhibit 
saturation and thresholding at, respectively, large and low 
values of total synaptic input. 
The corresponding Cauchy network generically exhibits two phases: 
quiescent and active, and an associated transition between 
them \cite{SI}. 
In general the nature of the active phase will depend on the details 
of the activation function.

To further simplify the calculations and to 
simultaneously model edge of chaos and avalanches, 
in the following we focus our attention on the binary activation function 
${\phi(x) = \Theta(x - \theta)}$, 
where $\Theta$ denotes the Heaviside function and 
$\theta$ denotes the threshold.
In this case the mean-field equation 
(\ref{eq:general-sc-mf}) simplifies to 
\begin{equation}
    m_{t+1} 
    = 
    \frac{1}{\pi}\arctan\left( m_t g /\theta \right).
    \label{eq:mf-binary}
\end{equation}
The stability of the trivial fixed point 
can be checked by expanding the RHS of 
(\ref{eq:mf-binary}) around $m_t = 0$:
$m_{t+1} = \frac{g}{\pi \theta}m_t + O(m_t^3)$.
The fixed point at $m_t = 0$, corresponding to the quiescent phase, 
is unstable for $g > \pi \theta$. 
Since $\arctan(x)/\pi$ is saturating and concave for all $x>0$,
another stable fixed point $m^*$ close to $0$ 
appears, through the supercritical pitchfork bifurcation, 
exactly when the trivial fixed point loses its stability 
($m^*\approx \sqrt{3}(g/\theta)^{-3/2}\sqrt{(g/\theta)-\pi}$ 
near the transition point).
Due to the quenched, asymmetric disorder of the connectivity matrix we can expect 
this fixed point to represent a chaotic attractor of the network
\footnote{
In the binary case the system has a finite number of states for any finite $N$ 
and the attractor has to be periodic. 
Thus, formally, irregular aperiodic behavior can only be observed in the limit of $N\to\infty$. 
Chaos-like signatures can nonetheless be observed for finite $N$, 
e.g. the typical lengths of transients and cycles rapidly change around the predicted transition point, 
and grow exponentially with $N$ in the ``chaotic'' phase 
\cite{vreeswijk1998chaotic,luque2000lyapunov}.
}, 
with a large sensitivity to small perturbations. 
Our computer simulations confirm this prediction 
\cite{SI}. 

%%%% FIGURE 2 %%%%%
\begin{figure}
    \centering
    \includegraphics[trim={1.7cm 0.8cm 2.2cm 1.1cm},clip, width=1\linewidth]{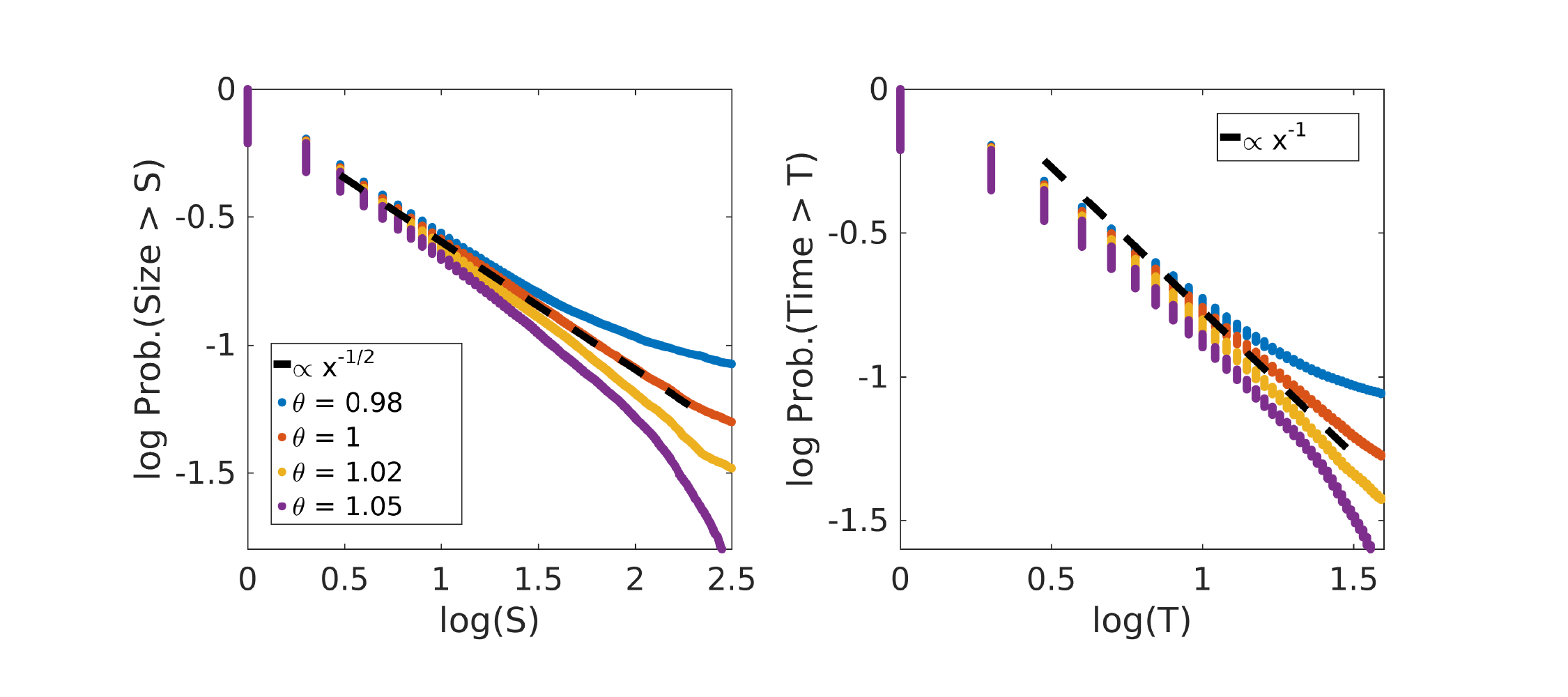}
    \caption{Avalanche size and lifetime distributions in the
    networks of binary units. 
    As expected from our theoretical predictions, 
    at the transition point these distributions are 
    described by power-laws, 
    and the critical exponents match those of the 
    critical branching process. 
    }
    \label{fig:avalanches}
\end{figure}

The transition from the quiescent to the chaotic phase 
can be understood from the underlying structure of connections. 
Due to the power-law connectivity density, 
we can expect that only a small fraction of the connections 
contribute to the activity profile of the network. 
Indeed, as we show in the following, the transition to chaos is driven by 
the percolation transition of {\it autocrat} connections for which ${J_{ij}>\theta}$, 
i.e. an active pre-synaptic neuron will activate the post-synaptic 
neuron in the absence of other inputs. 
Around the critical point the mean activity of the network is 
infinitesimal and thus the higher order interaction events 
(e.g. two neurons activating another neuron) are negligible. 
In other words, to a good approximation, a neuron can only be activated by another 
single neuron through an autocrat connection, independently from other neurons. 
This suggests that the transition to chaos in the neural network model 
is related to the critical branching processes \cite{harris2002theory} (Fig.~\ref{fig:nn}).

%%%% FIGURE 3 %%%%%
\begin{figure*}
    \centering
        \includegraphics[width= .34\linewidth]{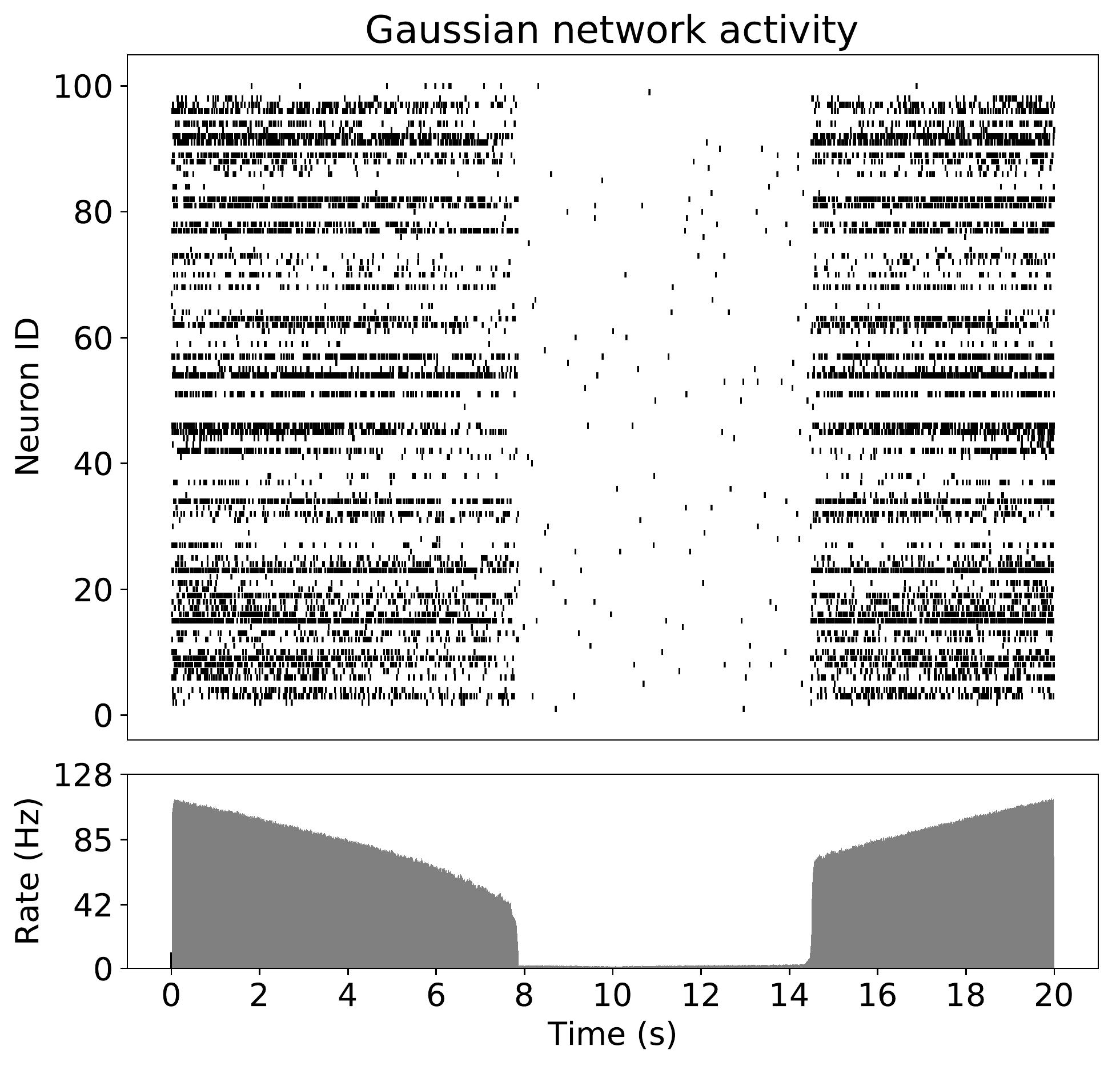}
        \includegraphics[width =.34\linewidth]{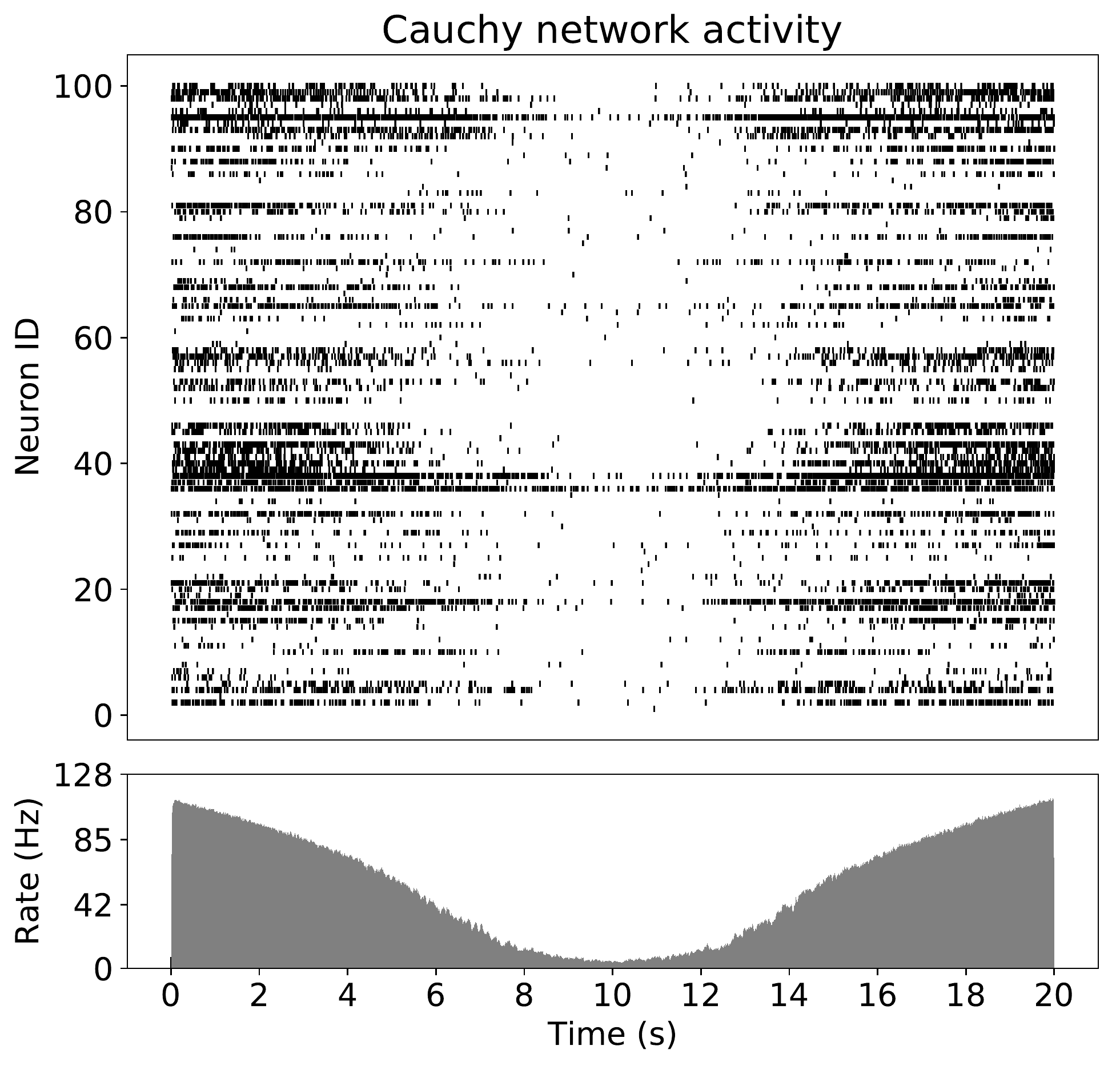}
        \includegraphics[width= .31\linewidth]{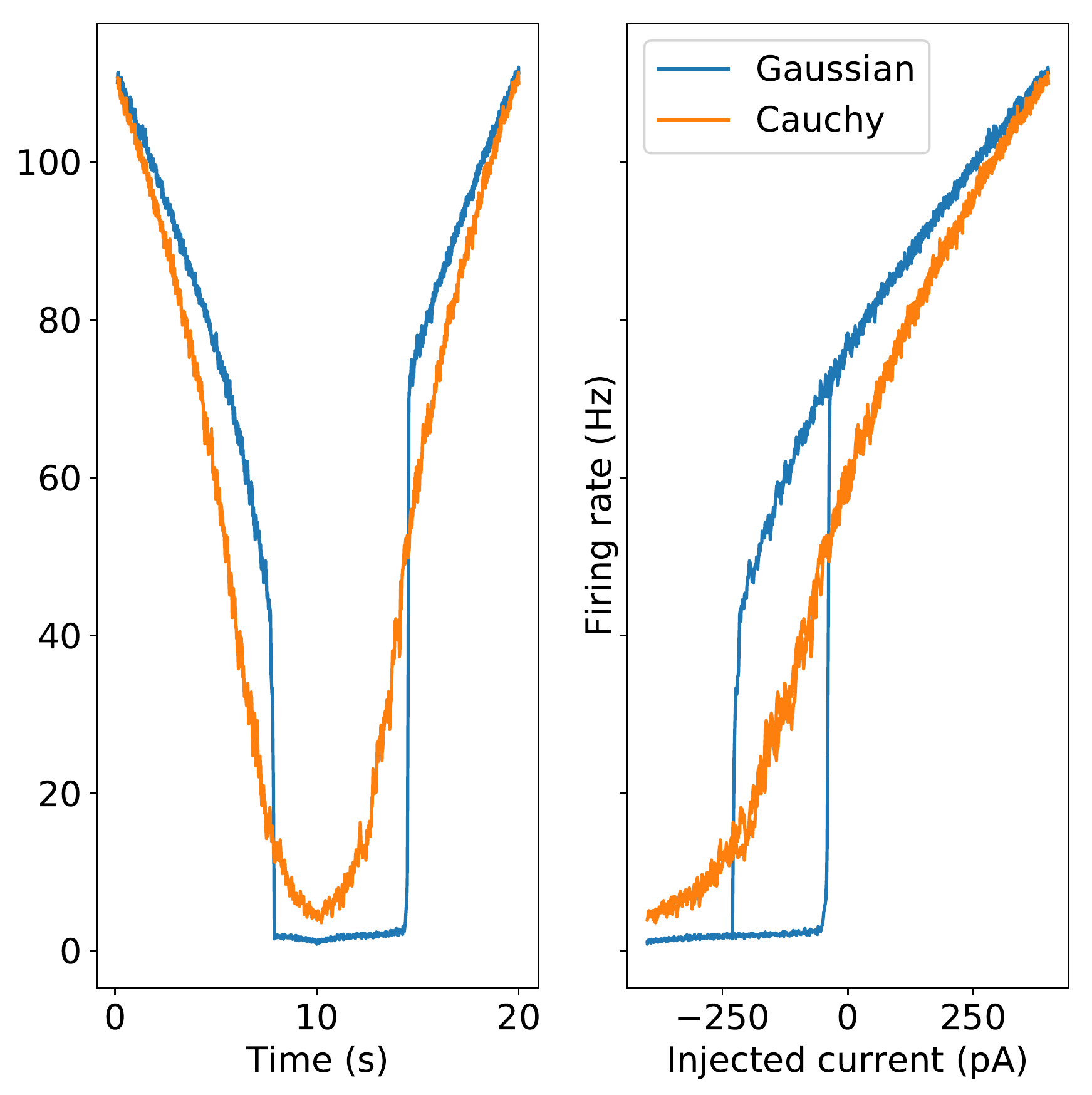}
        \caption{Continuous vs. discontinuous transition in networks of leaky integrate-and-fire neurons: 
    A slowly changing current was injected and an average firing rate of the network 
    was recorded as a function of time and the injected current amplitude. 
    As predicted by our theory, a network with Gaussian weights exhibits a 
    discontinuous transition between active and inactive states, 
    which generates a characteristic hysteresis loop. 
    In contrast, the Cauchy network exhibits a continuous 
    transition and thus shows no signs of the hysteresis loop.
    }
    \label{fig:spiking}
\end{figure*}
% =====================================================================================================
In the Cauchy case 
the probability that a given connection is an autocrat reads
\begin{equation}
    \Prob(J_{ij} > \theta) = 
    \frac{1}{\pi}\arctan\left(\frac{g}{N\theta}\right).
    \label{eq:success-prob}
\end{equation}
For a given neuron, the number of outgoing (or incoming) autocrat connections
is a binomial random variable with $N$ trials and the probability of success 
given by
(\ref{eq:success-prob}). 
In the limit of $N\to\infty$ it converges to the Poisson random variable 
with intensity 
\begin{equation}
    \lambda
    = 
    \lim\limits_{N\to\infty} \frac{N}{\pi}\arctan\left(\frac{g}{N\theta}\right) 
    =
    \frac{g}{\theta\pi}.
    \label{eq:lambda}
\end{equation}
Now, let the initial state of the network be such that only a single neuron 
(seed) is active. 
The number of active neurons (descendants) in the next step is 
given by the Poisson distribution and the mean number of active neurons 
is given by (\ref{eq:lambda}). 
The theory of branching processes predicts that the population will eventually
die out almost surely for $\lambda \leq 1$ and has a finite
survival probability for $\lambda > 1$.  
At $\lambda = 1$ the process is critical and features scale-free avalanches.
The critical point predicted by the branching process formulation 
of the network dynamics, $g^* = \pi \theta$, 
is the same as the mean-field critical point predicted by (\ref{eq:mf-binary}). 

The mapping to the branching process explains many features of 
the Cauchy neural network around the critical point. 
Below the critical point the steady state is quiescent and 
a bit-flip perturbation corresponds to a single neuron (seed) being activated. 
The local expansion rate of such perturbation is given by 
$\lambda$. 
Above the critical point ($g>\pi \theta$) 
each bit-flip contributes in the same manner as a single seed and, 
additionally, interacts with other active neurons to activate 
and deactivate other descendants. 
Thus, in the vicinity of the transition point
$\lambda$ gives a lower bound on the local expansion rate of 
a perturbation in the steady state, 
and for $\lambda>1$ the network is expected 
to be chaotic in the thermodynamic limit 
\cite{SI}. 
Moreover, the transition to chaos belongs to the 
mean-field directed percolation universality class 
\cite{alstrom1988mean, munoz1999avalanche, odor2004universality}. 
The propagation of the corresponding avalanches is characterized 
by \cite{SI} power-law distributed sizes $S$: 
$\Prob(S>s)\sim s^{-1/2}$, 
and power-law distributed lifetimes $T$: 
$\Prob(T>t)\sim t^{-1}$.
These theoretical predictions were corroborated by our 
computer simulations of the Cauchy network, as shown in 
Fig.~\ref{fig:avalanches}.

For a comparison, we have also studied Gaussian networks of threshold units with 
a fixed number of connections per neuron $K$ \cite{SI, derrida1987exactly}. 
While extremely sparsely connected Gaussian networks 
($K \lessapprox 12$) 
behave qualitatively similar to the Cauchy network, 
the transition to chaos becomes discontinuous in the biologically 
relevant regime of $K \gtrapprox 13$. 
With a biologically realistic $K$ and finite $N$, 
the network activity jumps between two metastable states 
near the transition point, and cannot be robustly posed at the edge of chaos. 
The discontinuous property is due to the emergence of 
a metastable active state by the saddle-node bifurcation 
as $g/\theta$ increases.

Importantly, although our theoretical predictions 
were derived assuming simplistic threshold neural units, 
they translate directly to networks of more biologically
plausible leaky integrate-and-fire (LIF) neurons.
The difference of continuous and discontinuous transition is 
confirmed by the presence or absence of a hysteresis loop in 
more realistic networks of LIF neurons
(Fig.~\ref{fig:spiking}). 
Hence, unlike the Gaussian networks with realistic $K$, 
Cauchy networks demonstrate critical phenomena and can reproduce 
experimentally observed scale-free avalanches at the critical point. 
Moreover, a large Cauchy network can exhibit arbitrarily low,
self-sustained activity levels. 
In contrast, the lowest possible activity level that can 
be achieved by the Gaussian network with realistic $K$ is 
about $11\%$ in the binary case and $40Hz$ in the LIF case (Fig.~\ref{fig:spiking}). 

For clarity we chose to limit our presentation to the Cauchy 
distribution of $J_{ij}$, 
but our results naturally extend to other power-law 
distributions.  
Indeed, let the synaptic efficacy density asymptotically behave 
like a power-law 
${\rho(J_{ij}) \sim C \alpha g^{\alpha} N^{-1} |J_{ij}|^{-1-\alpha}}$ 
\footnote{
Note that we scale $J_{ij}$ with the number of neurons as
$N^{-1/\alpha}$, 
which assures the existence of 
a non-trivial limit $N\to\infty$. 
For $\alpha>2$ another choice is to scale the synaptic 
strengths as $N^{-1/2}$, which in the limit of $N\to\infty$ 
corresponds to the Gaussian network. 
}.
We then have 
${\Prob(J_{ij} > \theta) = C  N^{-1} (g/\theta)^{\alpha}}$, 
which holds for large enough $N$. 
The branching parameter is calculated as in (\ref{eq:lambda}) 
and reads $\lambda=C (g/\theta)^{\alpha}$. 
A continuous transition takes place at $\lambda=1$ 
and its features are, as before, described by the 
directed percolation universality class. 

The connectivity in the current model is unstructured. 
It would be interesting to combine power-law synaptic weight distributions 
with structured networks, for example, with hierarchical modules 
\cite{friedman2013hierarchical} 
or oscillations
\cite{poil2012critical,di2018landau,fontenele2019criticality}.

Incidentally, the same Cauchy distribution of membrane potential 
was found in the quadratic integrate and fire neuron model 
\cite{montbrio2015macroscopic} due to single-neuron dynamics. 
Nontrivial effects may arise from a combination of 
single-neuron- and network-driven heavy-tail statistics. 

Power-law distributions of synaptic weights 
feature many very weak synapses, 
that do not directly contribute to the computation. 
Even though this may seem wasteful, we think that such architectures 
are not only biologically plausible \cite{cossell2015functional} 
but may be beneficial. 
One possibility is that even weak connections can activate a neuron once contextual 
input from another part of the brain increases the baseline membrane potential close 
to its spiking threshold. 
Such contextual input can also raise the spiking probability of nearby neurons 
so that synchronous activation of weak connections is more likely.
Weak synapses have also been reported to play a role in 
unsupervised features extraction \cite{huang2018role}. 
In the context of reservoir computing 
\cite{jaeger2001echo,maass2002real} 
high computational capabilities 
were achieved using non-biologically sparse connectivity
\cite{busing2010connectivity}. 
Our model provides a more biologically plausible solution that 
the connectivity can be anatomically dense but effectively 
sparse due to heavy-tailed synaptic weight distribution.  
More generally, the optimal degree of sparsity 
depends on the role of a given brain structure and the type of the employed plasticity 
\cite{litwin2017optimal}. 
Power-law distributed synaptic weights may in this context provide 
a weakly informative \cite{gelman2008weakly,van2006prior} sparse-connectivity prior, 
with weak and effectively silent synapses providing a pool of potential connections 
that can be recruited when and if needed, as observed in the brain during development 
\cite{liao1995activation,kerchner2008silent}. 

Our results demonstrate that the shape of the synaptic weight
distribution can dramatically affect dynamics of neural networks. 
A biological distribution of synaptic weights can give distinct predictions 
from the frequently assumed Gaussian distribution.
The proposed mathematical framework with power-law synaptic weights 
can easily be adapted to other scenarios in future studies.

%%%%%%%%%%%%%%%%%%%%%%%%%%%%%%%%%%%%%%%%%%%%%%%%%%%%%%%%%%%%%%%%%%%%%
\begin{acknowledgments}
We thank Francesco Fumarola and anonymous referees for invaluable
discussions and comments on the manuscript. 
Supported by RIKEN Center for Brain Science, Brain/MINDS from AMED under Grant Number
JP20dm020700, and JSPS KAKENHI Grant Number JP18H05432.
\end{acknowledgments}

%%%%%%%%%%%%%%%%%%%%%%%%%%%%%%%%%%%%%%%%%%%%%%%%%%%%%%%%%%%%%%%%%%%%%

\bibliography{references}

\begin{thebibliography}{82}%
\makeatletter
\providecommand \@ifxundefined [1]{%
 \@ifx{#1\undefined}
}%
\providecommand \@ifnum [1]{%
 \ifnum #1\expandafter \@firstoftwo
 \else \expandafter \@secondoftwo
 \fi
}%
\providecommand \@ifx [1]{%
 \ifx #1\expandafter \@firstoftwo
 \else \expandafter \@secondoftwo
 \fi
}%
\providecommand \natexlab [1]{#1}%
\providecommand \enquote  [1]{``#1''}%
\providecommand \bibnamefont  [1]{#1}%
\providecommand \bibfnamefont [1]{#1}%
\providecommand \citenamefont [1]{#1}%
\providecommand \href@noop [0]{\@secondoftwo}%
\providecommand \href [0]{\begingroup \@sanitize@url \@href}%
\providecommand \@href[1]{\@@startlink{#1}\@@href}%
\providecommand \@@href[1]{\endgroup#1\@@endlink}%
\providecommand \@sanitize@url [0]{\catcode `\\12\catcode `\$12\catcode
  `\&12\catcode `\#12\catcode `\^12\catcode `\_12\catcode `\%12\relax}%
\providecommand \@@startlink[1]{}%
\providecommand \@@endlink[0]{}%
\providecommand \url  [0]{\begingroup\@sanitize@url \@url }%
\providecommand \@url [1]{\endgroup\@href {#1}{\urlprefix }}%
\providecommand \urlprefix  [0]{URL }%
\providecommand \Eprint [0]{\href }%
\providecommand \doibase [0]{http://dx.doi.org/}%
\providecommand \selectlanguage [0]{\@gobble}%
\providecommand \bibinfo  [0]{\@secondoftwo}%
\providecommand \bibfield  [0]{\@secondoftwo}%
\providecommand \translation [1]{[#1]}%
\providecommand \BibitemOpen [0]{}%
\providecommand \bibitemStop [0]{}%
\providecommand \bibitemNoStop [0]{.\EOS\space}%
\providecommand \EOS [0]{\spacefactor3000\relax}%
\providecommand \BibitemShut  [1]{\csname bibitem#1\endcsname}%
\let\auto@bib@innerbib\@empty
%</preamble>
\bibitem [{\citenamefont {Beggs}\ and\ \citenamefont
  {Plenz}(2003)}]{beggs2003neuronal}%
  \BibitemOpen
  \bibfield  {author} {\bibinfo {author} {\bibfnamefont {J.~M.}\ \bibnamefont
  {Beggs}}\ and\ \bibinfo {author} {\bibfnamefont {D.}~\bibnamefont {Plenz}},\
  }\href@noop {} {\bibfield  {journal} {\bibinfo  {journal} {Journal of
  neuroscience}\ }\textbf {\bibinfo {volume} {23}},\ \bibinfo {pages} {11167}
  (\bibinfo {year} {2003})}\BibitemShut {NoStop}%
\bibitem [{\citenamefont {Shew}\ \emph {et~al.}(2015)\citenamefont {Shew},
  \citenamefont {Clawson}, \citenamefont {Pobst}, \citenamefont {Karimipanah},
  \citenamefont {Wright},\ and\ \citenamefont {Wessel}}]{shew2015adaptation}%
  \BibitemOpen
  \bibfield  {author} {\bibinfo {author} {\bibfnamefont {W.~L.}\ \bibnamefont
  {Shew}}, \bibinfo {author} {\bibfnamefont {W.~P.}\ \bibnamefont {Clawson}},
  \bibinfo {author} {\bibfnamefont {J.}~\bibnamefont {Pobst}}, \bibinfo
  {author} {\bibfnamefont {Y.}~\bibnamefont {Karimipanah}}, \bibinfo {author}
  {\bibfnamefont {N.~C.}\ \bibnamefont {Wright}}, \ and\ \bibinfo {author}
  {\bibfnamefont {R.}~\bibnamefont {Wessel}},\ }\href@noop {} {\bibfield
  {journal} {\bibinfo  {journal} {Nature Physics}\ }\textbf {\bibinfo {volume}
  {11}},\ \bibinfo {pages} {659} (\bibinfo {year} {2015})}\BibitemShut
  {NoStop}%
\bibitem [{\citenamefont {Fontenele}\ \emph {et~al.}(2019)\citenamefont
  {Fontenele}, \citenamefont {de~Vasconcelos}, \citenamefont {Feliciano},
  \citenamefont {Aguiar}, \citenamefont {Soares-Cunha}, \citenamefont
  {Coimbra}, \citenamefont {Dalla~Porta}, \citenamefont {Ribeiro},
  \citenamefont {Rodrigues}, \citenamefont {Sousa} \emph
  {et~al.}}]{fontenele2019criticality}%
  \BibitemOpen
  \bibfield  {author} {\bibinfo {author} {\bibfnamefont {A.~J.}\ \bibnamefont
  {Fontenele}}, \bibinfo {author} {\bibfnamefont {N.~A.}\ \bibnamefont
  {de~Vasconcelos}}, \bibinfo {author} {\bibfnamefont {T.}~\bibnamefont
  {Feliciano}}, \bibinfo {author} {\bibfnamefont {L.~A.}\ \bibnamefont
  {Aguiar}}, \bibinfo {author} {\bibfnamefont {C.}~\bibnamefont
  {Soares-Cunha}}, \bibinfo {author} {\bibfnamefont {B.}~\bibnamefont
  {Coimbra}}, \bibinfo {author} {\bibfnamefont {L.}~\bibnamefont
  {Dalla~Porta}}, \bibinfo {author} {\bibfnamefont {S.}~\bibnamefont
  {Ribeiro}}, \bibinfo {author} {\bibfnamefont {A.~J.}\ \bibnamefont
  {Rodrigues}}, \bibinfo {author} {\bibfnamefont {N.}~\bibnamefont {Sousa}},
  \emph {et~al.},\ }\href@noop {} {\bibfield  {journal} {\bibinfo  {journal}
  {Physical review letters}\ }\textbf {\bibinfo {volume} {122}},\ \bibinfo
  {pages} {208101} (\bibinfo {year} {2019})}\BibitemShut {NoStop}%
\bibitem [{\citenamefont {Petermann}\ \emph {et~al.}(2009)\citenamefont
  {Petermann}, \citenamefont {Thiagarajan}, \citenamefont {Lebedev},
  \citenamefont {Nicolelis}, \citenamefont {Chialvo},\ and\ \citenamefont
  {Plenz}}]{petermann2009spontaneous}%
  \BibitemOpen
  \bibfield  {author} {\bibinfo {author} {\bibfnamefont {T.}~\bibnamefont
  {Petermann}}, \bibinfo {author} {\bibfnamefont {T.~C.}\ \bibnamefont
  {Thiagarajan}}, \bibinfo {author} {\bibfnamefont {M.~A.}\ \bibnamefont
  {Lebedev}}, \bibinfo {author} {\bibfnamefont {M.~A.}\ \bibnamefont
  {Nicolelis}}, \bibinfo {author} {\bibfnamefont {D.~R.}\ \bibnamefont
  {Chialvo}}, \ and\ \bibinfo {author} {\bibfnamefont {D.}~\bibnamefont
  {Plenz}},\ }\href@noop {} {\bibfield  {journal} {\bibinfo  {journal}
  {Proceedings of the National Academy of Sciences}\ }\textbf {\bibinfo
  {volume} {106}},\ \bibinfo {pages} {15921} (\bibinfo {year}
  {2009})}\BibitemShut {NoStop}%
\bibitem [{\citenamefont {Haldeman}\ and\ \citenamefont
  {Beggs}(2005)}]{haldeman2005critical}%
  \BibitemOpen
  \bibfield  {author} {\bibinfo {author} {\bibfnamefont {C.}~\bibnamefont
  {Haldeman}}\ and\ \bibinfo {author} {\bibfnamefont {J.~M.}\ \bibnamefont
  {Beggs}},\ }\href@noop {} {\bibfield  {journal} {\bibinfo  {journal}
  {Physical review letters}\ }\textbf {\bibinfo {volume} {94}},\ \bibinfo
  {pages} {058101} (\bibinfo {year} {2005})}\BibitemShut {NoStop}%
\bibitem [{\citenamefont {Kinouchi}\ and\ \citenamefont
  {Copelli}(2006)}]{kinouchi2006optimal}%
  \BibitemOpen
  \bibfield  {author} {\bibinfo {author} {\bibfnamefont {O.}~\bibnamefont
  {Kinouchi}}\ and\ \bibinfo {author} {\bibfnamefont {M.}~\bibnamefont
  {Copelli}},\ }\href@noop {} {\bibfield  {journal} {\bibinfo  {journal}
  {Nature physics}\ }\textbf {\bibinfo {volume} {2}},\ \bibinfo {pages} {348}
  (\bibinfo {year} {2006})}\BibitemShut {NoStop}%
\bibitem [{\citenamefont {di~Santo}\ \emph {et~al.}(2018)\citenamefont
  {di~Santo}, \citenamefont {Villegas}, \citenamefont {Burioni},\ and\
  \citenamefont {Mu{\~n}oz}}]{di2018landau}%
  \BibitemOpen
  \bibfield  {author} {\bibinfo {author} {\bibfnamefont {S.}~\bibnamefont
  {di~Santo}}, \bibinfo {author} {\bibfnamefont {P.}~\bibnamefont {Villegas}},
  \bibinfo {author} {\bibfnamefont {R.}~\bibnamefont {Burioni}}, \ and\
  \bibinfo {author} {\bibfnamefont {M.~A.}\ \bibnamefont {Mu{\~n}oz}},\
  }\href@noop {} {\bibfield  {journal} {\bibinfo  {journal} {Proceedings of the
  National Academy of Sciences}\ }\textbf {\bibinfo {volume} {115}},\ \bibinfo
  {pages} {E1356} (\bibinfo {year} {2018})}\BibitemShut {NoStop}%
\bibitem [{\citenamefont {Stoop}\ and\ \citenamefont
  {Gomez}(2016)}]{stoop2016auditory}%
  \BibitemOpen
  \bibfield  {author} {\bibinfo {author} {\bibfnamefont {R.}~\bibnamefont
  {Stoop}}\ and\ \bibinfo {author} {\bibfnamefont {F.}~\bibnamefont {Gomez}},\
  }\href@noop {} {\bibfield  {journal} {\bibinfo  {journal} {Physical review
  letters}\ }\textbf {\bibinfo {volume} {117}},\ \bibinfo {pages} {038102}
  (\bibinfo {year} {2016})}\BibitemShut {NoStop}%
\bibitem [{\citenamefont {Munoz}(2018)}]{munoz2018colloquium}%
  \BibitemOpen
  \bibfield  {author} {\bibinfo {author} {\bibfnamefont {M.~A.}\ \bibnamefont
  {Munoz}},\ }\href@noop {} {\bibfield  {journal} {\bibinfo  {journal} {Reviews
  of Modern Physics}\ }\textbf {\bibinfo {volume} {90}},\ \bibinfo {pages}
  {031001} (\bibinfo {year} {2018})}\BibitemShut {NoStop}%
\bibitem [{\citenamefont {Langton}(1990)}]{langton1990computation}%
  \BibitemOpen
  \bibfield  {author} {\bibinfo {author} {\bibfnamefont {C.~G.}\ \bibnamefont
  {Langton}},\ }\href@noop {} {\bibfield  {journal} {\bibinfo  {journal}
  {Physica D: Nonlinear Phenomena}\ }\textbf {\bibinfo {volume} {42}},\
  \bibinfo {pages} {12} (\bibinfo {year} {1990})}\BibitemShut {NoStop}%
\bibitem [{\citenamefont {Bertschinger}\ and\ \citenamefont
  {Natschl{\"a}ger}(2004)}]{bertschinger2004real}%
  \BibitemOpen
  \bibfield  {author} {\bibinfo {author} {\bibfnamefont {N.}~\bibnamefont
  {Bertschinger}}\ and\ \bibinfo {author} {\bibfnamefont {T.}~\bibnamefont
  {Natschl{\"a}ger}},\ }\href@noop {} {\bibfield  {journal} {\bibinfo
  {journal} {Neural computation}\ }\textbf {\bibinfo {volume} {16}},\ \bibinfo
  {pages} {1413} (\bibinfo {year} {2004})}\BibitemShut {NoStop}%
\bibitem [{\citenamefont {Legenstein}\ and\ \citenamefont
  {Maass}(2007)}]{legenstein2007makes}%
  \BibitemOpen
  \bibfield  {author} {\bibinfo {author} {\bibfnamefont {R.}~\bibnamefont
  {Legenstein}}\ and\ \bibinfo {author} {\bibfnamefont {W.}~\bibnamefont
  {Maass}},\ }\href@noop {} {\bibfield  {journal} {\bibinfo  {journal} {New
  directions in statistical signal processing: From systems to brain}\ ,\
  \bibinfo {pages} {127}} (\bibinfo {year} {2007})}\BibitemShut {NoStop}%
\bibitem [{\citenamefont {Toyoizumi}\ and\ \citenamefont
  {Abbott}(2011)}]{toyoizumi2011beyond}%
  \BibitemOpen
  \bibfield  {author} {\bibinfo {author} {\bibfnamefont {T.}~\bibnamefont
  {Toyoizumi}}\ and\ \bibinfo {author} {\bibfnamefont {L.}~\bibnamefont
  {Abbott}},\ }\href@noop {} {\bibfield  {journal} {\bibinfo  {journal}
  {Physical Review E}\ }\textbf {\bibinfo {volume} {84}},\ \bibinfo {pages}
  {051908} (\bibinfo {year} {2011})}\BibitemShut {NoStop}%
\bibitem [{\citenamefont {Schuecker}\ \emph {et~al.}(2018)\citenamefont
  {Schuecker}, \citenamefont {Goedeke},\ and\ \citenamefont
  {Helias}}]{schuecker2018optimal}%
  \BibitemOpen
  \bibfield  {author} {\bibinfo {author} {\bibfnamefont {J.}~\bibnamefont
  {Schuecker}}, \bibinfo {author} {\bibfnamefont {S.}~\bibnamefont {Goedeke}},
  \ and\ \bibinfo {author} {\bibfnamefont {M.}~\bibnamefont {Helias}},\
  }\href@noop {} {\bibfield  {journal} {\bibinfo  {journal} {Physical Review
  X}\ }\textbf {\bibinfo {volume} {8}},\ \bibinfo {pages} {041029} (\bibinfo
  {year} {2018})}\BibitemShut {NoStop}%
\bibitem [{\citenamefont {Benayoun}\ \emph {et~al.}(2010)\citenamefont
  {Benayoun}, \citenamefont {Cowan}, \citenamefont {van Drongelen},\ and\
  \citenamefont {Wallace}}]{benayoun2010avalanches}%
  \BibitemOpen
  \bibfield  {author} {\bibinfo {author} {\bibfnamefont {M.}~\bibnamefont
  {Benayoun}}, \bibinfo {author} {\bibfnamefont {J.~D.}\ \bibnamefont {Cowan}},
  \bibinfo {author} {\bibfnamefont {W.}~\bibnamefont {van Drongelen}}, \ and\
  \bibinfo {author} {\bibfnamefont {E.}~\bibnamefont {Wallace}},\ }\href@noop
  {} {\bibfield  {journal} {\bibinfo  {journal} {PLoS computational biology}\
  }\textbf {\bibinfo {volume} {6}},\ \bibinfo {pages} {e1000846} (\bibinfo
  {year} {2010})}\BibitemShut {NoStop}%
\bibitem [{\citenamefont {Beggs}\ and\ \citenamefont
  {Timme}(2012)}]{beggs2012being}%
  \BibitemOpen
  \bibfield  {author} {\bibinfo {author} {\bibfnamefont {J.~M.}\ \bibnamefont
  {Beggs}}\ and\ \bibinfo {author} {\bibfnamefont {N.}~\bibnamefont {Timme}},\
  }\href@noop {} {\bibfield  {journal} {\bibinfo  {journal} {Frontiers in
  physiology}\ }\textbf {\bibinfo {volume} {3}},\ \bibinfo {pages} {163}
  (\bibinfo {year} {2012})}\BibitemShut {NoStop}%
\bibitem [{\citenamefont {Sompolinsky}\ \emph {et~al.}(1988)\citenamefont
  {Sompolinsky}, \citenamefont {Crisanti},\ and\ \citenamefont
  {Sommers}}]{sompolinsky1988chaos}%
  \BibitemOpen
  \bibfield  {author} {\bibinfo {author} {\bibfnamefont {H.}~\bibnamefont
  {Sompolinsky}}, \bibinfo {author} {\bibfnamefont {A.}~\bibnamefont
  {Crisanti}}, \ and\ \bibinfo {author} {\bibfnamefont {H.-J.}\ \bibnamefont
  {Sommers}},\ }\href@noop {} {\bibfield  {journal} {\bibinfo  {journal}
  {Physical review letters}\ }\textbf {\bibinfo {volume} {61}},\ \bibinfo
  {pages} {259} (\bibinfo {year} {1988})}\BibitemShut {NoStop}%
\bibitem [{\citenamefont {Molgedey}\ \emph {et~al.}(1992)\citenamefont
  {Molgedey}, \citenamefont {Schuchhardt},\ and\ \citenamefont
  {Schuster}}]{molgedey1992suppressing}%
  \BibitemOpen
  \bibfield  {author} {\bibinfo {author} {\bibfnamefont {L.}~\bibnamefont
  {Molgedey}}, \bibinfo {author} {\bibfnamefont {J.}~\bibnamefont
  {Schuchhardt}}, \ and\ \bibinfo {author} {\bibfnamefont {H.~G.}\ \bibnamefont
  {Schuster}},\ }\href@noop {} {\bibfield  {journal} {\bibinfo  {journal}
  {Physical review letters}\ }\textbf {\bibinfo {volume} {69}},\ \bibinfo
  {pages} {3717} (\bibinfo {year} {1992})}\BibitemShut {NoStop}%
\bibitem [{\citenamefont {Amit}\ and\ \citenamefont
  {Brunel}(1997)}]{amit1997model}%
  \BibitemOpen
  \bibfield  {author} {\bibinfo {author} {\bibfnamefont {D.~J.}\ \bibnamefont
  {Amit}}\ and\ \bibinfo {author} {\bibfnamefont {N.}~\bibnamefont {Brunel}},\
  }\href@noop {} {\bibfield  {journal} {\bibinfo  {journal} {Cerebral cortex
  (New York, NY: 1991)}\ }\textbf {\bibinfo {volume} {7}},\ \bibinfo {pages}
  {237} (\bibinfo {year} {1997})}\BibitemShut {NoStop}%
\bibitem [{\citenamefont {Brunel}(2000)}]{brunel2000dynamics}%
  \BibitemOpen
  \bibfield  {author} {\bibinfo {author} {\bibfnamefont {N.}~\bibnamefont
  {Brunel}},\ }\href@noop {} {\bibfield  {journal} {\bibinfo  {journal}
  {Journal of computational neuroscience}\ }\textbf {\bibinfo {volume} {8}},\
  \bibinfo {pages} {183} (\bibinfo {year} {2000})}\BibitemShut {NoStop}%
\bibitem [{\citenamefont {Rajan}\ \emph {et~al.}(2010)\citenamefont {Rajan},
  \citenamefont {Abbott},\ and\ \citenamefont
  {Sompolinsky}}]{rajan2010stimulus}%
  \BibitemOpen
  \bibfield  {author} {\bibinfo {author} {\bibfnamefont {K.}~\bibnamefont
  {Rajan}}, \bibinfo {author} {\bibfnamefont {L.}~\bibnamefont {Abbott}}, \
  and\ \bibinfo {author} {\bibfnamefont {H.}~\bibnamefont {Sompolinsky}},\
  }\href@noop {} {\bibfield  {journal} {\bibinfo  {journal} {Physical Review
  E}\ }\textbf {\bibinfo {volume} {82}},\ \bibinfo {pages} {011903} (\bibinfo
  {year} {2010})}\BibitemShut {NoStop}%
\bibitem [{\citenamefont {Massar}\ and\ \citenamefont
  {Massar}(2013)}]{massar2013mean}%
  \BibitemOpen
  \bibfield  {author} {\bibinfo {author} {\bibfnamefont {M.}~\bibnamefont
  {Massar}}\ and\ \bibinfo {author} {\bibfnamefont {S.}~\bibnamefont
  {Massar}},\ }\href@noop {} {\bibfield  {journal} {\bibinfo  {journal}
  {Physical Review E}\ }\textbf {\bibinfo {volume} {87}},\ \bibinfo {pages}
  {042809} (\bibinfo {year} {2013})}\BibitemShut {NoStop}%
\bibitem [{\citenamefont {Stern}\ \emph {et~al.}(2014)\citenamefont {Stern},
  \citenamefont {Sompolinsky},\ and\ \citenamefont
  {Abbott}}]{stern2014dynamics}%
  \BibitemOpen
  \bibfield  {author} {\bibinfo {author} {\bibfnamefont {M.}~\bibnamefont
  {Stern}}, \bibinfo {author} {\bibfnamefont {H.}~\bibnamefont {Sompolinsky}},
  \ and\ \bibinfo {author} {\bibfnamefont {L.}~\bibnamefont {Abbott}},\
  }\href@noop {} {\bibfield  {journal} {\bibinfo  {journal} {Physical Review
  E}\ }\textbf {\bibinfo {volume} {90}},\ \bibinfo {pages} {062710} (\bibinfo
  {year} {2014})}\BibitemShut {NoStop}%
\bibitem [{\citenamefont {Aljadeff}\ \emph {et~al.}(2015)\citenamefont
  {Aljadeff}, \citenamefont {Stern},\ and\ \citenamefont
  {Sharpee}}]{aljadeff2015transition}%
  \BibitemOpen
  \bibfield  {author} {\bibinfo {author} {\bibfnamefont {J.}~\bibnamefont
  {Aljadeff}}, \bibinfo {author} {\bibfnamefont {M.}~\bibnamefont {Stern}}, \
  and\ \bibinfo {author} {\bibfnamefont {T.}~\bibnamefont {Sharpee}},\
  }\href@noop {} {\bibfield  {journal} {\bibinfo  {journal} {Physical review
  letters}\ }\textbf {\bibinfo {volume} {114}},\ \bibinfo {pages} {088101}
  (\bibinfo {year} {2015})}\BibitemShut {NoStop}%
\bibitem [{\citenamefont {Kadmon}\ and\ \citenamefont
  {Sompolinsky}(2015)}]{kadmon2015transition}%
  \BibitemOpen
  \bibfield  {author} {\bibinfo {author} {\bibfnamefont {J.}~\bibnamefont
  {Kadmon}}\ and\ \bibinfo {author} {\bibfnamefont {H.}~\bibnamefont
  {Sompolinsky}},\ }\href@noop {} {\bibfield  {journal} {\bibinfo  {journal}
  {Physical Review X}\ }\textbf {\bibinfo {volume} {5}},\ \bibinfo {pages}
  {041030} (\bibinfo {year} {2015})}\BibitemShut {NoStop}%
\bibitem [{\citenamefont {Wainrib}\ and\ \citenamefont
  {Galtier}(2016)}]{wainrib2016local}%
  \BibitemOpen
  \bibfield  {author} {\bibinfo {author} {\bibfnamefont {G.}~\bibnamefont
  {Wainrib}}\ and\ \bibinfo {author} {\bibfnamefont {M.~N.}\ \bibnamefont
  {Galtier}},\ }\href@noop {} {\bibfield  {journal} {\bibinfo  {journal}
  {Neural Networks}\ }\textbf {\bibinfo {volume} {76}},\ \bibinfo {pages} {39}
  (\bibinfo {year} {2016})}\BibitemShut {NoStop}%
\bibitem [{\citenamefont {Crisanti}\ and\ \citenamefont
  {Sompolinsky}(2018)}]{crisanti2018path}%
  \BibitemOpen
  \bibfield  {author} {\bibinfo {author} {\bibfnamefont {A.}~\bibnamefont
  {Crisanti}}\ and\ \bibinfo {author} {\bibfnamefont {H.}~\bibnamefont
  {Sompolinsky}},\ }\href@noop {} {\bibfield  {journal} {\bibinfo  {journal}
  {Physical Review E}\ }\textbf {\bibinfo {volume} {98}},\ \bibinfo {pages}
  {062120} (\bibinfo {year} {2018})}\BibitemShut {NoStop}%
\bibitem [{\citenamefont {Harish}\ and\ \citenamefont
  {Hansel}(2015)}]{harish2015asynchronous}%
  \BibitemOpen
  \bibfield  {author} {\bibinfo {author} {\bibfnamefont {O.}~\bibnamefont
  {Harish}}\ and\ \bibinfo {author} {\bibfnamefont {D.}~\bibnamefont
  {Hansel}},\ }\href@noop {} {\bibfield  {journal} {\bibinfo  {journal} {PLoS
  computational biology}\ }\textbf {\bibinfo {volume} {11}} (\bibinfo {year}
  {2015})}\BibitemShut {NoStop}%
\bibitem [{\citenamefont {Dayan}\ and\ \citenamefont
  {Abbott}(2001)}]{dayan2001theoretical}%
  \BibitemOpen
  \bibfield  {author} {\bibinfo {author} {\bibfnamefont {P.}~\bibnamefont
  {Dayan}}\ and\ \bibinfo {author} {\bibfnamefont {L.~F.}\ \bibnamefont
  {Abbott}},\ }\href@noop {} {\  (\bibinfo {year} {2001})}\BibitemShut
  {NoStop}%
\bibitem [{\citenamefont {H{\"a}usser}\ and\ \citenamefont
  {Monsivais}(2003)}]{hausser2003less}%
  \BibitemOpen
  \bibfield  {author} {\bibinfo {author} {\bibfnamefont {M.}~\bibnamefont
  {H{\"a}usser}}\ and\ \bibinfo {author} {\bibfnamefont {P.}~\bibnamefont
  {Monsivais}},\ }\href@noop {} {\bibfield  {journal} {\bibinfo  {journal}
  {Neuron}\ }\textbf {\bibinfo {volume} {40}},\ \bibinfo {pages} {449}
  (\bibinfo {year} {2003})}\BibitemShut {NoStop}%
\bibitem [{\citenamefont {Gerstner}\ \emph {et~al.}(2014)\citenamefont
  {Gerstner}, \citenamefont {Kistler}, \citenamefont {Naud},\ and\
  \citenamefont {Paninski}}]{gerstner2014neuronal}%
  \BibitemOpen
  \bibfield  {author} {\bibinfo {author} {\bibfnamefont {W.}~\bibnamefont
  {Gerstner}}, \bibinfo {author} {\bibfnamefont {W.~M.}\ \bibnamefont
  {Kistler}}, \bibinfo {author} {\bibfnamefont {R.}~\bibnamefont {Naud}}, \
  and\ \bibinfo {author} {\bibfnamefont {L.}~\bibnamefont {Paninski}},\
  }\href@noop {} {\emph {\bibinfo {title} {Neuronal dynamics: From single
  neurons to networks and models of cognition}}}\ (\bibinfo  {publisher}
  {Cambridge University Press},\ \bibinfo {year} {2014})\BibitemShut {NoStop}%
\bibitem [{\citenamefont {Millman}\ \emph {et~al.}(2010)\citenamefont
  {Millman}, \citenamefont {Mihalas}, \citenamefont {Kirkwood},\ and\
  \citenamefont {Niebur}}]{millman2010self}%
  \BibitemOpen
  \bibfield  {author} {\bibinfo {author} {\bibfnamefont {D.}~\bibnamefont
  {Millman}}, \bibinfo {author} {\bibfnamefont {S.}~\bibnamefont {Mihalas}},
  \bibinfo {author} {\bibfnamefont {A.}~\bibnamefont {Kirkwood}}, \ and\
  \bibinfo {author} {\bibfnamefont {E.}~\bibnamefont {Niebur}},\ }\href@noop {}
  {\bibfield  {journal} {\bibinfo  {journal} {Nature physics}\ }\textbf
  {\bibinfo {volume} {6}},\ \bibinfo {pages} {801} (\bibinfo {year}
  {2010})}\BibitemShut {NoStop}%
\bibitem [{\citenamefont {Dayan}\ and\ \citenamefont
  {Abbott}(2005)}]{dayan2005book}%
  \BibitemOpen
  \bibfield  {author} {\bibinfo {author} {\bibfnamefont {P.}~\bibnamefont
  {Dayan}}\ and\ \bibinfo {author} {\bibfnamefont {L.~F.}\ \bibnamefont
  {Abbott}},\ }\href@noop {} {\emph {\bibinfo {title} {Theoretical
  Neuroscience: Computational and Mathematical Modeling of Neural Systems}}}\
  (\bibinfo  {publisher} {The MIT Press},\ \bibinfo {year} {2005})\BibitemShut
  {NoStop}%
\bibitem [{\citenamefont {Marshel}\ \emph {et~al.}(2019)\citenamefont
  {Marshel}, \citenamefont {Kim}, \citenamefont {Machado}, \citenamefont
  {Quirin}, \citenamefont {Benson}, \citenamefont {Kadmon}, \citenamefont
  {Raja}, \citenamefont {Chibukhchyan}, \citenamefont {Ramakrishnan},
  \citenamefont {Inoue} \emph {et~al.}}]{marshel2019cortical}%
  \BibitemOpen
  \bibfield  {author} {\bibinfo {author} {\bibfnamefont {J.~H.}\ \bibnamefont
  {Marshel}}, \bibinfo {author} {\bibfnamefont {Y.~S.}\ \bibnamefont {Kim}},
  \bibinfo {author} {\bibfnamefont {T.~A.}\ \bibnamefont {Machado}}, \bibinfo
  {author} {\bibfnamefont {S.}~\bibnamefont {Quirin}}, \bibinfo {author}
  {\bibfnamefont {B.}~\bibnamefont {Benson}}, \bibinfo {author} {\bibfnamefont
  {J.}~\bibnamefont {Kadmon}}, \bibinfo {author} {\bibfnamefont
  {C.}~\bibnamefont {Raja}}, \bibinfo {author} {\bibfnamefont {A.}~\bibnamefont
  {Chibukhchyan}}, \bibinfo {author} {\bibfnamefont {C.}~\bibnamefont
  {Ramakrishnan}}, \bibinfo {author} {\bibfnamefont {M.}~\bibnamefont {Inoue}},
   \emph {et~al.},\ }\href@noop {} {\bibfield  {journal} {\bibinfo  {journal}
  {Science}\ }\textbf {\bibinfo {volume} {365}},\ \bibinfo {pages} {eaaw5202}
  (\bibinfo {year} {2019})}\BibitemShut {NoStop}%
\bibitem [{\citenamefont {Scarpetta}\ \emph {et~al.}(2018)\citenamefont
  {Scarpetta}, \citenamefont {Apicella}, \citenamefont {Minati},\ and\
  \citenamefont {de~Candia}}]{scarpetta2018hysteresis}%
  \BibitemOpen
  \bibfield  {author} {\bibinfo {author} {\bibfnamefont {S.}~\bibnamefont
  {Scarpetta}}, \bibinfo {author} {\bibfnamefont {I.}~\bibnamefont {Apicella}},
  \bibinfo {author} {\bibfnamefont {L.}~\bibnamefont {Minati}}, \ and\ \bibinfo
  {author} {\bibfnamefont {A.}~\bibnamefont {de~Candia}},\ }\href@noop {}
  {\bibfield  {journal} {\bibinfo  {journal} {Physical Review E}\ }\textbf
  {\bibinfo {volume} {97}},\ \bibinfo {pages} {062305} (\bibinfo {year}
  {2018})}\BibitemShut {NoStop}%
\bibitem [{\citenamefont {Sayer}\ \emph {et~al.}(1990)\citenamefont {Sayer},
  \citenamefont {Friedlander},\ and\ \citenamefont {Redman}}]{sayer1990time}%
  \BibitemOpen
  \bibfield  {author} {\bibinfo {author} {\bibfnamefont {R.}~\bibnamefont
  {Sayer}}, \bibinfo {author} {\bibfnamefont {M.}~\bibnamefont {Friedlander}},
  \ and\ \bibinfo {author} {\bibfnamefont {S.}~\bibnamefont {Redman}},\
  }\href@noop {} {\bibfield  {journal} {\bibinfo  {journal} {Journal of
  Neuroscience}\ }\textbf {\bibinfo {volume} {10}},\ \bibinfo {pages} {826}
  (\bibinfo {year} {1990})}\BibitemShut {NoStop}%
\bibitem [{\citenamefont {Feldmeyer}\ \emph {et~al.}(1999)\citenamefont
  {Feldmeyer}, \citenamefont {Egger}, \citenamefont {L{\"u}bke},\ and\
  \citenamefont {Sakmann}}]{feldmeyer1999reliable}%
  \BibitemOpen
  \bibfield  {author} {\bibinfo {author} {\bibfnamefont {D.}~\bibnamefont
  {Feldmeyer}}, \bibinfo {author} {\bibfnamefont {V.}~\bibnamefont {Egger}},
  \bibinfo {author} {\bibfnamefont {J.}~\bibnamefont {L{\"u}bke}}, \ and\
  \bibinfo {author} {\bibfnamefont {B.}~\bibnamefont {Sakmann}},\ }\href@noop
  {} {\bibfield  {journal} {\bibinfo  {journal} {The Journal of physiology}\
  }\textbf {\bibinfo {volume} {521}},\ \bibinfo {pages} {169} (\bibinfo {year}
  {1999})}\BibitemShut {NoStop}%
\bibitem [{\citenamefont {Song}\ \emph {et~al.}(2005)\citenamefont {Song},
  \citenamefont {Sj{\"o}str{\"o}m}, \citenamefont {Reigl}, \citenamefont
  {Nelson},\ and\ \citenamefont {Chklovskii}}]{song2005highly}%
  \BibitemOpen
  \bibfield  {author} {\bibinfo {author} {\bibfnamefont {S.}~\bibnamefont
  {Song}}, \bibinfo {author} {\bibfnamefont {P.~J.}\ \bibnamefont
  {Sj{\"o}str{\"o}m}}, \bibinfo {author} {\bibfnamefont {M.}~\bibnamefont
  {Reigl}}, \bibinfo {author} {\bibfnamefont {S.}~\bibnamefont {Nelson}}, \
  and\ \bibinfo {author} {\bibfnamefont {D.~B.}\ \bibnamefont {Chklovskii}},\
  }\href@noop {} {\bibfield  {journal} {\bibinfo  {journal} {PLoS biology}\
  }\textbf {\bibinfo {volume} {3}},\ \bibinfo {pages} {e68} (\bibinfo {year}
  {2005})}\BibitemShut {NoStop}%
\bibitem [{\citenamefont {Lefort}\ \emph {et~al.}(2009)\citenamefont {Lefort},
  \citenamefont {Tomm}, \citenamefont {Sarria},\ and\ \citenamefont
  {Petersen}}]{lefort2009excitatory}%
  \BibitemOpen
  \bibfield  {author} {\bibinfo {author} {\bibfnamefont {S.}~\bibnamefont
  {Lefort}}, \bibinfo {author} {\bibfnamefont {C.}~\bibnamefont {Tomm}},
  \bibinfo {author} {\bibfnamefont {J.-C.~F.}\ \bibnamefont {Sarria}}, \ and\
  \bibinfo {author} {\bibfnamefont {C.~C.}\ \bibnamefont {Petersen}},\
  }\href@noop {} {\bibfield  {journal} {\bibinfo  {journal} {Neuron}\ }\textbf
  {\bibinfo {volume} {61}},\ \bibinfo {pages} {301} (\bibinfo {year}
  {2009})}\BibitemShut {NoStop}%
\bibitem [{\citenamefont {Ikegaya}\ \emph {et~al.}(2012)\citenamefont
  {Ikegaya}, \citenamefont {Sasaki}, \citenamefont {Ishikawa}, \citenamefont
  {Honma}, \citenamefont {Tao}, \citenamefont {Takahashi}, \citenamefont
  {Minamisawa}, \citenamefont {Ujita},\ and\ \citenamefont
  {Matsuki}}]{ikegaya2012interpyramid}%
  \BibitemOpen
  \bibfield  {author} {\bibinfo {author} {\bibfnamefont {Y.}~\bibnamefont
  {Ikegaya}}, \bibinfo {author} {\bibfnamefont {T.}~\bibnamefont {Sasaki}},
  \bibinfo {author} {\bibfnamefont {D.}~\bibnamefont {Ishikawa}}, \bibinfo
  {author} {\bibfnamefont {N.}~\bibnamefont {Honma}}, \bibinfo {author}
  {\bibfnamefont {K.}~\bibnamefont {Tao}}, \bibinfo {author} {\bibfnamefont
  {N.}~\bibnamefont {Takahashi}}, \bibinfo {author} {\bibfnamefont
  {G.}~\bibnamefont {Minamisawa}}, \bibinfo {author} {\bibfnamefont
  {S.}~\bibnamefont {Ujita}}, \ and\ \bibinfo {author} {\bibfnamefont
  {N.}~\bibnamefont {Matsuki}},\ }\href@noop {} {\bibfield  {journal} {\bibinfo
   {journal} {Cerebral Cortex}\ }\textbf {\bibinfo {volume} {23}},\ \bibinfo
  {pages} {293} (\bibinfo {year} {2012})}\BibitemShut {NoStop}%
\bibitem [{\citenamefont {Loewenstein}\ \emph {et~al.}(2011)\citenamefont
  {Loewenstein}, \citenamefont {Kuras},\ and\ \citenamefont
  {Rumpel}}]{loewenstein2011multiplicative}%
  \BibitemOpen
  \bibfield  {author} {\bibinfo {author} {\bibfnamefont {Y.}~\bibnamefont
  {Loewenstein}}, \bibinfo {author} {\bibfnamefont {A.}~\bibnamefont {Kuras}},
  \ and\ \bibinfo {author} {\bibfnamefont {S.}~\bibnamefont {Rumpel}},\
  }\href@noop {} {\bibfield  {journal} {\bibinfo  {journal} {Journal of
  Neuroscience}\ }\textbf {\bibinfo {volume} {31}},\ \bibinfo {pages} {9481}
  (\bibinfo {year} {2011})}\BibitemShut {NoStop}%
\bibitem [{\citenamefont {Gilson}\ and\ \citenamefont
  {Fukai}(2011)}]{gilson2011stability}%
  \BibitemOpen
  \bibfield  {author} {\bibinfo {author} {\bibfnamefont {M.}~\bibnamefont
  {Gilson}}\ and\ \bibinfo {author} {\bibfnamefont {T.}~\bibnamefont {Fukai}},\
  }\href@noop {} {\bibfield  {journal} {\bibinfo  {journal} {PloS one}\
  }\textbf {\bibinfo {volume} {6}},\ \bibinfo {pages} {e25339} (\bibinfo {year}
  {2011})}\BibitemShut {NoStop}%
\bibitem [{\citenamefont {Zheng}\ \emph {et~al.}(2013)\citenamefont {Zheng},
  \citenamefont {Dimitrakakis},\ and\ \citenamefont
  {Triesch}}]{zheng2013network}%
  \BibitemOpen
  \bibfield  {author} {\bibinfo {author} {\bibfnamefont {P.}~\bibnamefont
  {Zheng}}, \bibinfo {author} {\bibfnamefont {C.}~\bibnamefont {Dimitrakakis}},
  \ and\ \bibinfo {author} {\bibfnamefont {J.}~\bibnamefont {Triesch}},\
  }\href@noop {} {\bibfield  {journal} {\bibinfo  {journal} {PLoS computational
  biology}\ }\textbf {\bibinfo {volume} {9}},\ \bibinfo {pages} {e1002848}
  (\bibinfo {year} {2013})}\BibitemShut {NoStop}%
\bibitem [{\citenamefont {Yasumatsu}\ \emph {et~al.}(2008)\citenamefont
  {Yasumatsu}, \citenamefont {Matsuzaki}, \citenamefont {Miyazaki},
  \citenamefont {Noguchi},\ and\ \citenamefont
  {Kasai}}]{yasumatsu2008principles}%
  \BibitemOpen
  \bibfield  {author} {\bibinfo {author} {\bibfnamefont {N.}~\bibnamefont
  {Yasumatsu}}, \bibinfo {author} {\bibfnamefont {M.}~\bibnamefont
  {Matsuzaki}}, \bibinfo {author} {\bibfnamefont {T.}~\bibnamefont {Miyazaki}},
  \bibinfo {author} {\bibfnamefont {J.}~\bibnamefont {Noguchi}}, \ and\
  \bibinfo {author} {\bibfnamefont {H.}~\bibnamefont {Kasai}},\ }\href@noop {}
  {\bibfield  {journal} {\bibinfo  {journal} {Journal of Neuroscience}\
  }\textbf {\bibinfo {volume} {28}},\ \bibinfo {pages} {13592} (\bibinfo {year}
  {2008})}\BibitemShut {NoStop}%
\bibitem [{\citenamefont {Nagaoka}\ \emph {et~al.}(2016)\citenamefont
  {Nagaoka}, \citenamefont {Takehara}, \citenamefont {Hayashi-Takagi},
  \citenamefont {Noguchi}, \citenamefont {Ishii}, \citenamefont {Shirai},
  \citenamefont {Yagishita}, \citenamefont {Akagi}, \citenamefont {Ichiki},\
  and\ \citenamefont {Kasai}}]{nagaoka2016abnormal}%
  \BibitemOpen
  \bibfield  {author} {\bibinfo {author} {\bibfnamefont {A.}~\bibnamefont
  {Nagaoka}}, \bibinfo {author} {\bibfnamefont {H.}~\bibnamefont {Takehara}},
  \bibinfo {author} {\bibfnamefont {A.}~\bibnamefont {Hayashi-Takagi}},
  \bibinfo {author} {\bibfnamefont {J.}~\bibnamefont {Noguchi}}, \bibinfo
  {author} {\bibfnamefont {K.}~\bibnamefont {Ishii}}, \bibinfo {author}
  {\bibfnamefont {F.}~\bibnamefont {Shirai}}, \bibinfo {author} {\bibfnamefont
  {S.}~\bibnamefont {Yagishita}}, \bibinfo {author} {\bibfnamefont
  {T.}~\bibnamefont {Akagi}}, \bibinfo {author} {\bibfnamefont
  {T.}~\bibnamefont {Ichiki}}, \ and\ \bibinfo {author} {\bibfnamefont
  {H.}~\bibnamefont {Kasai}},\ }\href@noop {} {\bibfield  {journal} {\bibinfo
  {journal} {Scientific reports}\ }\textbf {\bibinfo {volume} {6}},\ \bibinfo
  {pages} {26651} (\bibinfo {year} {2016})}\BibitemShut {NoStop}%
\bibitem [{\citenamefont {Humble}\ \emph {et~al.}(2019)\citenamefont {Humble},
  \citenamefont {Hiratsuka}, \citenamefont {Kasai},\ and\ \citenamefont
  {Toyoizumi}}]{humble2019intrinsic}%
  \BibitemOpen
  \bibfield  {author} {\bibinfo {author} {\bibfnamefont {J.}~\bibnamefont
  {Humble}}, \bibinfo {author} {\bibfnamefont {K.}~\bibnamefont {Hiratsuka}},
  \bibinfo {author} {\bibfnamefont {H.}~\bibnamefont {Kasai}}, \ and\ \bibinfo
  {author} {\bibfnamefont {T.}~\bibnamefont {Toyoizumi}},\ }\href@noop {}
  {\bibfield  {journal} {\bibinfo  {journal} {Frontiers in Computational
  Neuroscience}\ }\textbf {\bibinfo {volume} {13}},\ \bibinfo {pages} {38}
  (\bibinfo {year} {2019})}\BibitemShut {NoStop}%
\bibitem [{\citenamefont {Ishii}\ \emph {et~al.}(2018)\citenamefont {Ishii},
  \citenamefont {Nagaoka}, \citenamefont {Kishida}, \citenamefont {Okazaki},
  \citenamefont {Yagishita}, \citenamefont {Ucar}, \citenamefont {Takahashi},
  \citenamefont {Saito},\ and\ \citenamefont {Kasai}}]{ishii2018vivo}%
  \BibitemOpen
  \bibfield  {author} {\bibinfo {author} {\bibfnamefont {K.}~\bibnamefont
  {Ishii}}, \bibinfo {author} {\bibfnamefont {A.}~\bibnamefont {Nagaoka}},
  \bibinfo {author} {\bibfnamefont {Y.}~\bibnamefont {Kishida}}, \bibinfo
  {author} {\bibfnamefont {H.}~\bibnamefont {Okazaki}}, \bibinfo {author}
  {\bibfnamefont {S.}~\bibnamefont {Yagishita}}, \bibinfo {author}
  {\bibfnamefont {H.}~\bibnamefont {Ucar}}, \bibinfo {author} {\bibfnamefont
  {N.}~\bibnamefont {Takahashi}}, \bibinfo {author} {\bibfnamefont
  {N.}~\bibnamefont {Saito}}, \ and\ \bibinfo {author} {\bibfnamefont
  {H.}~\bibnamefont {Kasai}},\ }\href@noop {} {\bibfield  {journal} {\bibinfo
  {journal} {eNeuro}\ }\textbf {\bibinfo {volume} {5}} (\bibinfo {year}
  {2018})}\BibitemShut {NoStop}%
\bibitem [{\citenamefont {Okazaki}\ \emph {et~al.}(2018)\citenamefont
  {Okazaki}, \citenamefont {Hayashi-Takagi}, \citenamefont {Nagaoka},
  \citenamefont {Negishi}, \citenamefont {Ucar}, \citenamefont {Yagishita},
  \citenamefont {Ishii}, \citenamefont {Toyoizumi}, \citenamefont {Fox},\ and\
  \citenamefont {Kasai}}]{okazaki2018calcineurin}%
  \BibitemOpen
  \bibfield  {author} {\bibinfo {author} {\bibfnamefont {H.}~\bibnamefont
  {Okazaki}}, \bibinfo {author} {\bibfnamefont {A.}~\bibnamefont
  {Hayashi-Takagi}}, \bibinfo {author} {\bibfnamefont {A.}~\bibnamefont
  {Nagaoka}}, \bibinfo {author} {\bibfnamefont {M.}~\bibnamefont {Negishi}},
  \bibinfo {author} {\bibfnamefont {H.}~\bibnamefont {Ucar}}, \bibinfo {author}
  {\bibfnamefont {S.}~\bibnamefont {Yagishita}}, \bibinfo {author}
  {\bibfnamefont {K.}~\bibnamefont {Ishii}}, \bibinfo {author} {\bibfnamefont
  {T.}~\bibnamefont {Toyoizumi}}, \bibinfo {author} {\bibfnamefont
  {K.}~\bibnamefont {Fox}}, \ and\ \bibinfo {author} {\bibfnamefont
  {H.}~\bibnamefont {Kasai}},\ }\href@noop {} {\bibfield  {journal} {\bibinfo
  {journal} {Neuroscience letters}\ }\textbf {\bibinfo {volume} {671}},\
  \bibinfo {pages} {99} (\bibinfo {year} {2018})}\BibitemShut {NoStop}%
\bibitem [{\citenamefont {Teramae}\ \emph {et~al.}(2012)\citenamefont
  {Teramae}, \citenamefont {Tsubo},\ and\ \citenamefont
  {Fukai}}]{teramae2012optimal}%
  \BibitemOpen
  \bibfield  {author} {\bibinfo {author} {\bibfnamefont {J.-n.}\ \bibnamefont
  {Teramae}}, \bibinfo {author} {\bibfnamefont {Y.}~\bibnamefont {Tsubo}}, \
  and\ \bibinfo {author} {\bibfnamefont {T.}~\bibnamefont {Fukai}},\
  }\href@noop {} {\bibfield  {journal} {\bibinfo  {journal} {Scientific
  reports}\ }\textbf {\bibinfo {volume} {2}},\ \bibinfo {pages} {485} (\bibinfo
  {year} {2012})}\BibitemShut {NoStop}%
\bibitem [{\citenamefont {Buzs{\'a}ki}\ and\ \citenamefont
  {Mizuseki}(2014)}]{buzsaki2014log}%
  \BibitemOpen
  \bibfield  {author} {\bibinfo {author} {\bibfnamefont {G.}~\bibnamefont
  {Buzs{\'a}ki}}\ and\ \bibinfo {author} {\bibfnamefont {K.}~\bibnamefont
  {Mizuseki}},\ }\href@noop {} {\bibfield  {journal} {\bibinfo  {journal}
  {Nature Reviews Neuroscience}\ }\textbf {\bibinfo {volume} {15}},\ \bibinfo
  {pages} {264} (\bibinfo {year} {2014})}\BibitemShut {NoStop}%
\bibitem [{\citenamefont {Teramae}\ and\ \citenamefont
  {Fukai}(2014)}]{teramae2014computational}%
  \BibitemOpen
  \bibfield  {author} {\bibinfo {author} {\bibfnamefont {J.-n.}\ \bibnamefont
  {Teramae}}\ and\ \bibinfo {author} {\bibfnamefont {T.}~\bibnamefont
  {Fukai}},\ }\href@noop {} {\bibfield  {journal} {\bibinfo  {journal}
  {Proceedings of the IEEE}\ }\textbf {\bibinfo {volume} {102}},\ \bibinfo
  {pages} {500} (\bibinfo {year} {2014})}\BibitemShut {NoStop}%
\bibitem [{\citenamefont {Omura}\ \emph {et~al.}(2015)\citenamefont {Omura},
  \citenamefont {Carvalho}, \citenamefont {Inokuchi},\ and\ \citenamefont
  {Fukai}}]{omura2015lognormal}%
  \BibitemOpen
  \bibfield  {author} {\bibinfo {author} {\bibfnamefont {Y.}~\bibnamefont
  {Omura}}, \bibinfo {author} {\bibfnamefont {M.~M.}\ \bibnamefont {Carvalho}},
  \bibinfo {author} {\bibfnamefont {K.}~\bibnamefont {Inokuchi}}, \ and\
  \bibinfo {author} {\bibfnamefont {T.}~\bibnamefont {Fukai}},\ }\href@noop {}
  {\bibfield  {journal} {\bibinfo  {journal} {Journal of Neuroscience}\
  }\textbf {\bibinfo {volume} {35}},\ \bibinfo {pages} {14585} (\bibinfo {year}
  {2015})}\BibitemShut {NoStop}%
\bibitem [{\citenamefont {Feller}(1971)}]{feller1971introduction}%
  \BibitemOpen
  \bibfield  {author} {\bibinfo {author} {\bibfnamefont {W.}~\bibnamefont
  {Feller}},\ }\href@noop {} {\emph {\bibinfo {title} {An introduction to
  probability and its applications, Vol. II}}}\ (\bibinfo  {publisher} {Wiley,
  New York},\ \bibinfo {year} {1971})\BibitemShut {NoStop}%
\bibitem [{\citenamefont {Cizeau}\ and\ \citenamefont
  {Bouchaud}(1994)}]{cizeau1994theory}%
  \BibitemOpen
  \bibfield  {author} {\bibinfo {author} {\bibfnamefont {P.}~\bibnamefont
  {Cizeau}}\ and\ \bibinfo {author} {\bibfnamefont {J.-P.}\ \bibnamefont
  {Bouchaud}},\ }\href@noop {} {\bibfield  {journal} {\bibinfo  {journal}
  {Physical Review E}\ }\textbf {\bibinfo {volume} {50}},\ \bibinfo {pages}
  {1810} (\bibinfo {year} {1994})}\BibitemShut {NoStop}%
\bibitem [{\citenamefont {Burda}\ \emph {et~al.}(2002)\citenamefont {Burda},
  \citenamefont {Janik}, \citenamefont {Jurkiewicz}, \citenamefont {Nowak},
  \citenamefont {Papp},\ and\ \citenamefont {Zahed}}]{burda2002free}%
  \BibitemOpen
  \bibfield  {author} {\bibinfo {author} {\bibfnamefont {Z.}~\bibnamefont
  {Burda}}, \bibinfo {author} {\bibfnamefont {R.~A.}\ \bibnamefont {Janik}},
  \bibinfo {author} {\bibfnamefont {J.}~\bibnamefont {Jurkiewicz}}, \bibinfo
  {author} {\bibfnamefont {M.~A.}\ \bibnamefont {Nowak}}, \bibinfo {author}
  {\bibfnamefont {G.}~\bibnamefont {Papp}}, \ and\ \bibinfo {author}
  {\bibfnamefont {I.}~\bibnamefont {Zahed}},\ }\href@noop {} {\bibfield
  {journal} {\bibinfo  {journal} {Physical Review E}\ }\textbf {\bibinfo
  {volume} {65}},\ \bibinfo {pages} {021106} (\bibinfo {year}
  {2002})}\BibitemShut {NoStop}%
\bibitem [{\citenamefont {Gudowska-Nowak}\ \emph {et~al.}(2020)\citenamefont
  {Gudowska-Nowak}, \citenamefont {Nowak}, \citenamefont {Chialvo},
  \citenamefont {Ochab},\ and\ \citenamefont
  {Tarnowski}}]{gudowska2020synaptic}%
  \BibitemOpen
  \bibfield  {author} {\bibinfo {author} {\bibfnamefont {E.}~\bibnamefont
  {Gudowska-Nowak}}, \bibinfo {author} {\bibfnamefont {M.~A.}\ \bibnamefont
  {Nowak}}, \bibinfo {author} {\bibfnamefont {D.~R.}\ \bibnamefont {Chialvo}},
  \bibinfo {author} {\bibfnamefont {J.~K.}\ \bibnamefont {Ochab}}, \ and\
  \bibinfo {author} {\bibfnamefont {W.}~\bibnamefont {Tarnowski}},\ }\href@noop
  {} {\bibfield  {journal} {\bibinfo  {journal} {Neural Computation}\ }\textbf
  {\bibinfo {volume} {32}},\ \bibinfo {pages} {395} (\bibinfo {year}
  {2020})}\BibitemShut {NoStop}%
\bibitem [{SI()}]{SI}%
  \BibitemOpen
  \href@noop {} {}\bibinfo {note} {See Supplementary Material for details,
  which includes
  Refs.~\cite{goles1986antisymmetrical,bezanson2017julia,nest216}.}\BibitemShut
  {Stop}%
\bibitem [{Note1()}]{Note1}%
  \BibitemOpen
  \bibinfo {note} {In the binary case the system has a finite number of states
  for any finite $N$ and the attractor has to be periodic. Thus, formally,
  irregular aperiodic behavior can only be observed in the limit of $N\to
  \infty $. Chaos-like signatures can nonetheless be observed for finite $N$,
  e.g. the typical lengths of transients and cycles rapidly change around the
  predicted transition point, and grow exponentially with $N$ in the
  ``chaotic'' phase \cite
  {vreeswijk1998chaotic,luque2000lyapunov}.}\BibitemShut {Stop}%
\bibitem [{\citenamefont {Harris}(2002)}]{harris2002theory}%
  \BibitemOpen
  \bibfield  {author} {\bibinfo {author} {\bibfnamefont {T.~E.}\ \bibnamefont
  {Harris}},\ }\href@noop {} {\emph {\bibinfo {title} {The theory of branching
  processes}}}\ (\bibinfo  {publisher} {Courier Corporation},\ \bibinfo {year}
  {2002})\BibitemShut {NoStop}%
\bibitem [{\citenamefont {Alstr{\o}m}(1988)}]{alstrom1988mean}%
  \BibitemOpen
  \bibfield  {author} {\bibinfo {author} {\bibfnamefont {P.}~\bibnamefont
  {Alstr{\o}m}},\ }\href@noop {} {\bibfield  {journal} {\bibinfo  {journal}
  {Physical Review A}\ }\textbf {\bibinfo {volume} {38}},\ \bibinfo {pages}
  {4905} (\bibinfo {year} {1988})}\BibitemShut {NoStop}%
\bibitem [{\citenamefont {Munoz}\ \emph {et~al.}(1999)\citenamefont {Munoz},
  \citenamefont {Dickman}, \citenamefont {Vespignani},\ and\ \citenamefont
  {Zapperi}}]{munoz1999avalanche}%
  \BibitemOpen
  \bibfield  {author} {\bibinfo {author} {\bibfnamefont {M.~A.}\ \bibnamefont
  {Munoz}}, \bibinfo {author} {\bibfnamefont {R.}~\bibnamefont {Dickman}},
  \bibinfo {author} {\bibfnamefont {A.}~\bibnamefont {Vespignani}}, \ and\
  \bibinfo {author} {\bibfnamefont {S.}~\bibnamefont {Zapperi}},\ }\href@noop
  {} {\bibfield  {journal} {\bibinfo  {journal} {Physical Review E}\ }\textbf
  {\bibinfo {volume} {59}},\ \bibinfo {pages} {6175} (\bibinfo {year}
  {1999})}\BibitemShut {NoStop}%
\bibitem [{\citenamefont {{\'O}dor}(2004)}]{odor2004universality}%
  \BibitemOpen
  \bibfield  {author} {\bibinfo {author} {\bibfnamefont {G.}~\bibnamefont
  {{\'O}dor}},\ }\href@noop {} {\bibfield  {journal} {\bibinfo  {journal}
  {Reviews of modern physics}\ }\textbf {\bibinfo {volume} {76}},\ \bibinfo
  {pages} {663} (\bibinfo {year} {2004})}\BibitemShut {NoStop}%
\bibitem [{\citenamefont {Derrida}\ \emph {et~al.}(1987)\citenamefont
  {Derrida}, \citenamefont {Gardner},\ and\ \citenamefont
  {Zippelius}}]{derrida1987exactly}%
  \BibitemOpen
  \bibfield  {author} {\bibinfo {author} {\bibfnamefont {B.}~\bibnamefont
  {Derrida}}, \bibinfo {author} {\bibfnamefont {E.}~\bibnamefont {Gardner}}, \
  and\ \bibinfo {author} {\bibfnamefont {A.}~\bibnamefont {Zippelius}},\
  }\href@noop {} {\bibfield  {journal} {\bibinfo  {journal} {EPL (Europhysics
  Letters)}\ }\textbf {\bibinfo {volume} {4}},\ \bibinfo {pages} {167}
  (\bibinfo {year} {1987})}\BibitemShut {NoStop}%
\bibitem [{Note2()}]{Note2}%
  \BibitemOpen
  \bibinfo {note} {Note that we scale $J_{ij}$ with the number of neurons as
  $N^{-1/\alpha }$, which assures the existence of a non-trivial limit $N\to
  \infty $. For $\alpha >2$ another choice is to scale the synaptic strengths
  as $N^{-1/2}$, which in the limit of $N\to \infty $ corresponds to the
  Gaussian network.}\BibitemShut {Stop}%
\bibitem [{\citenamefont {Friedman}\ and\ \citenamefont
  {Landsberg}(2013)}]{friedman2013hierarchical}%
  \BibitemOpen
  \bibfield  {author} {\bibinfo {author} {\bibfnamefont {E.~J.}\ \bibnamefont
  {Friedman}}\ and\ \bibinfo {author} {\bibfnamefont {A.~S.}\ \bibnamefont
  {Landsberg}},\ }\href@noop {} {\bibfield  {journal} {\bibinfo  {journal}
  {Chaos: An Interdisciplinary Journal of Nonlinear Science}\ }\textbf
  {\bibinfo {volume} {23}},\ \bibinfo {pages} {013135} (\bibinfo {year}
  {2013})}\BibitemShut {NoStop}%
\bibitem [{\citenamefont {Poil}\ \emph {et~al.}(2012)\citenamefont {Poil},
  \citenamefont {Hardstone}, \citenamefont {Mansvelder},\ and\ \citenamefont
  {Linkenkaer-Hansen}}]{poil2012critical}%
  \BibitemOpen
  \bibfield  {author} {\bibinfo {author} {\bibfnamefont {S.-S.}\ \bibnamefont
  {Poil}}, \bibinfo {author} {\bibfnamefont {R.}~\bibnamefont {Hardstone}},
  \bibinfo {author} {\bibfnamefont {H.~D.}\ \bibnamefont {Mansvelder}}, \ and\
  \bibinfo {author} {\bibfnamefont {K.}~\bibnamefont {Linkenkaer-Hansen}},\
  }\href@noop {} {\bibfield  {journal} {\bibinfo  {journal} {Journal of
  Neuroscience}\ }\textbf {\bibinfo {volume} {32}},\ \bibinfo {pages} {9817}
  (\bibinfo {year} {2012})}\BibitemShut {NoStop}%
\bibitem [{\citenamefont {Montbri{\'o}}\ \emph {et~al.}(2015)\citenamefont
  {Montbri{\'o}}, \citenamefont {Paz{\'o}},\ and\ \citenamefont
  {Roxin}}]{montbrio2015macroscopic}%
  \BibitemOpen
  \bibfield  {author} {\bibinfo {author} {\bibfnamefont {E.}~\bibnamefont
  {Montbri{\'o}}}, \bibinfo {author} {\bibfnamefont {D.}~\bibnamefont
  {Paz{\'o}}}, \ and\ \bibinfo {author} {\bibfnamefont {A.}~\bibnamefont
  {Roxin}},\ }\href@noop {} {\bibfield  {journal} {\bibinfo  {journal}
  {Physical Review X}\ }\textbf {\bibinfo {volume} {5}},\ \bibinfo {pages}
  {021028} (\bibinfo {year} {2015})}\BibitemShut {NoStop}%
\bibitem [{\citenamefont {Cossell}\ \emph {et~al.}(2015)\citenamefont
  {Cossell}, \citenamefont {Iacaruso}, \citenamefont {Muir}, \citenamefont
  {Houlton}, \citenamefont {Sader}, \citenamefont {Ko}, \citenamefont {Hofer},\
  and\ \citenamefont {Mrsic-Flogel}}]{cossell2015functional}%
  \BibitemOpen
  \bibfield  {author} {\bibinfo {author} {\bibfnamefont {L.}~\bibnamefont
  {Cossell}}, \bibinfo {author} {\bibfnamefont {M.~F.}\ \bibnamefont
  {Iacaruso}}, \bibinfo {author} {\bibfnamefont {D.~R.}\ \bibnamefont {Muir}},
  \bibinfo {author} {\bibfnamefont {R.}~\bibnamefont {Houlton}}, \bibinfo
  {author} {\bibfnamefont {E.~N.}\ \bibnamefont {Sader}}, \bibinfo {author}
  {\bibfnamefont {H.}~\bibnamefont {Ko}}, \bibinfo {author} {\bibfnamefont
  {S.~B.}\ \bibnamefont {Hofer}}, \ and\ \bibinfo {author} {\bibfnamefont
  {T.~D.}\ \bibnamefont {Mrsic-Flogel}},\ }\href@noop {} {\bibfield  {journal}
  {\bibinfo  {journal} {Nature}\ }\textbf {\bibinfo {volume} {518}},\ \bibinfo
  {pages} {399} (\bibinfo {year} {2015})}\BibitemShut {NoStop}%
\bibitem [{\citenamefont {Huang}(2018)}]{huang2018role}%
  \BibitemOpen
  \bibfield  {author} {\bibinfo {author} {\bibfnamefont {H.}~\bibnamefont
  {Huang}},\ }\href@noop {} {\bibfield  {journal} {\bibinfo  {journal} {Journal
  of Physics A: Mathematical and Theoretical}\ }\textbf {\bibinfo {volume}
  {51}},\ \bibinfo {pages} {08LT01} (\bibinfo {year} {2018})}\BibitemShut
  {NoStop}%
\bibitem [{\citenamefont {Jaeger}(2001)}]{jaeger2001echo}%
  \BibitemOpen
  \bibfield  {author} {\bibinfo {author} {\bibfnamefont {H.}~\bibnamefont
  {Jaeger}},\ }\href@noop {} {\bibfield  {journal} {\bibinfo  {journal} {Bonn,
  Germany: German National Research Center for Information Technology GMD
  Technical Report}\ }\textbf {\bibinfo {volume} {148}},\ \bibinfo {pages} {13}
  (\bibinfo {year} {2001})}\BibitemShut {NoStop}%
\bibitem [{\citenamefont {Maass}\ \emph {et~al.}(2002)\citenamefont {Maass},
  \citenamefont {Natschl{\"a}ger},\ and\ \citenamefont
  {Markram}}]{maass2002real}%
  \BibitemOpen
  \bibfield  {author} {\bibinfo {author} {\bibfnamefont {W.}~\bibnamefont
  {Maass}}, \bibinfo {author} {\bibfnamefont {T.}~\bibnamefont
  {Natschl{\"a}ger}}, \ and\ \bibinfo {author} {\bibfnamefont {H.}~\bibnamefont
  {Markram}},\ }\href@noop {} {\bibfield  {journal} {\bibinfo  {journal}
  {Neural computation}\ }\textbf {\bibinfo {volume} {14}},\ \bibinfo {pages}
  {2531} (\bibinfo {year} {2002})}\BibitemShut {NoStop}%
\bibitem [{\citenamefont {B{\"u}sing}\ \emph {et~al.}(2010)\citenamefont
  {B{\"u}sing}, \citenamefont {Schrauwen},\ and\ \citenamefont
  {Legenstein}}]{busing2010connectivity}%
  \BibitemOpen
  \bibfield  {author} {\bibinfo {author} {\bibfnamefont {L.}~\bibnamefont
  {B{\"u}sing}}, \bibinfo {author} {\bibfnamefont {B.}~\bibnamefont
  {Schrauwen}}, \ and\ \bibinfo {author} {\bibfnamefont {R.}~\bibnamefont
  {Legenstein}},\ }\href@noop {} {\bibfield  {journal} {\bibinfo  {journal}
  {Neural computation}\ }\textbf {\bibinfo {volume} {22}},\ \bibinfo {pages}
  {1272} (\bibinfo {year} {2010})}\BibitemShut {NoStop}%
\bibitem [{\citenamefont {Litwin-Kumar}\ \emph {et~al.}(2017)\citenamefont
  {Litwin-Kumar}, \citenamefont {Harris}, \citenamefont {Axel}, \citenamefont
  {Sompolinsky},\ and\ \citenamefont {Abbott}}]{litwin2017optimal}%
  \BibitemOpen
  \bibfield  {author} {\bibinfo {author} {\bibfnamefont {A.}~\bibnamefont
  {Litwin-Kumar}}, \bibinfo {author} {\bibfnamefont {K.~D.}\ \bibnamefont
  {Harris}}, \bibinfo {author} {\bibfnamefont {R.}~\bibnamefont {Axel}},
  \bibinfo {author} {\bibfnamefont {H.}~\bibnamefont {Sompolinsky}}, \ and\
  \bibinfo {author} {\bibfnamefont {L.}~\bibnamefont {Abbott}},\ }\href@noop {}
  {\bibfield  {journal} {\bibinfo  {journal} {Neuron}\ }\textbf {\bibinfo
  {volume} {93}},\ \bibinfo {pages} {1153} (\bibinfo {year}
  {2017})}\BibitemShut {NoStop}%
\bibitem [{\citenamefont {Gelman}\ \emph {et~al.}(2008)\citenamefont {Gelman},
  \citenamefont {Jakulin}, \citenamefont {Pittau}, \citenamefont {Su} \emph
  {et~al.}}]{gelman2008weakly}%
  \BibitemOpen
  \bibfield  {author} {\bibinfo {author} {\bibfnamefont {A.}~\bibnamefont
  {Gelman}}, \bibinfo {author} {\bibfnamefont {A.}~\bibnamefont {Jakulin}},
  \bibinfo {author} {\bibfnamefont {M.~G.}\ \bibnamefont {Pittau}}, \bibinfo
  {author} {\bibfnamefont {Y.-S.}\ \bibnamefont {Su}},  \emph {et~al.},\
  }\href@noop {} {\bibfield  {journal} {\bibinfo  {journal} {The Annals of
  Applied Statistics}\ }\textbf {\bibinfo {volume} {2}},\ \bibinfo {pages}
  {1360} (\bibinfo {year} {2008})}\BibitemShut {NoStop}%
\bibitem [{\citenamefont {Van~Dongen}(2006)}]{van2006prior}%
  \BibitemOpen
  \bibfield  {author} {\bibinfo {author} {\bibfnamefont {S.}~\bibnamefont
  {Van~Dongen}},\ }\href@noop {} {\bibfield  {journal} {\bibinfo  {journal}
  {Journal of Theoretical Biology}\ }\textbf {\bibinfo {volume} {242}},\
  \bibinfo {pages} {90} (\bibinfo {year} {2006})}\BibitemShut {NoStop}%
\bibitem [{\citenamefont {Liao}\ \emph {et~al.}(1995)\citenamefont {Liao},
  \citenamefont {Hessler},\ and\ \citenamefont {Malinow}}]{liao1995activation}%
  \BibitemOpen
  \bibfield  {author} {\bibinfo {author} {\bibfnamefont {D.}~\bibnamefont
  {Liao}}, \bibinfo {author} {\bibfnamefont {N.~A.}\ \bibnamefont {Hessler}}, \
  and\ \bibinfo {author} {\bibfnamefont {R.}~\bibnamefont {Malinow}},\
  }\href@noop {} {\bibfield  {journal} {\bibinfo  {journal} {Nature}\ }\textbf
  {\bibinfo {volume} {375}},\ \bibinfo {pages} {400} (\bibinfo {year}
  {1995})}\BibitemShut {NoStop}%
\bibitem [{\citenamefont {Kerchner}\ and\ \citenamefont
  {Nicoll}(2008)}]{kerchner2008silent}%
  \BibitemOpen
  \bibfield  {author} {\bibinfo {author} {\bibfnamefont {G.~A.}\ \bibnamefont
  {Kerchner}}\ and\ \bibinfo {author} {\bibfnamefont {R.~A.}\ \bibnamefont
  {Nicoll}},\ }\href@noop {} {\bibfield  {journal} {\bibinfo  {journal} {Nature
  Reviews Neuroscience}\ }\textbf {\bibinfo {volume} {9}},\ \bibinfo {pages}
  {813} (\bibinfo {year} {2008})}\BibitemShut {NoStop}%
\bibitem [{\citenamefont {Goles}(1986)}]{goles1986antisymmetrical}%
  \BibitemOpen
  \bibfield  {author} {\bibinfo {author} {\bibfnamefont {E.}~\bibnamefont
  {Goles}},\ }\href@noop {} {\bibfield  {journal} {\bibinfo  {journal}
  {Discrete Applied Mathematics}\ }\textbf {\bibinfo {volume} {13}},\ \bibinfo
  {pages} {97} (\bibinfo {year} {1986})}\BibitemShut {NoStop}%
\bibitem [{\citenamefont {Bezanson}\ \emph {et~al.}(2017)\citenamefont
  {Bezanson}, \citenamefont {Edelman}, \citenamefont {Karpinski},\ and\
  \citenamefont {Shah}}]{bezanson2017julia}%
  \BibitemOpen
  \bibfield  {author} {\bibinfo {author} {\bibfnamefont {J.}~\bibnamefont
  {Bezanson}}, \bibinfo {author} {\bibfnamefont {A.}~\bibnamefont {Edelman}},
  \bibinfo {author} {\bibfnamefont {S.}~\bibnamefont {Karpinski}}, \ and\
  \bibinfo {author} {\bibfnamefont {V.~B.}\ \bibnamefont {Shah}},\ }\href
  {https://doi.org/10.1137/141000671} {\bibfield  {journal} {\bibinfo
  {journal} {SIAM review}\ }\textbf {\bibinfo {volume} {59}},\ \bibinfo {pages}
  {65} (\bibinfo {year} {2017})}\BibitemShut {NoStop}%
\bibitem [{\citenamefont {Linssen}\ \emph {et~al.}(2018)\citenamefont
  {Linssen}, \citenamefont {Lepperød}, \citenamefont {Mitchell}, \citenamefont
  {Pronold}, \citenamefont {Eppler}, \citenamefont {Keup}, \citenamefont
  {Peyser}, \citenamefont {Kunkel}, \citenamefont {Weidel}, \citenamefont
  {Nodem}, \citenamefont {Terhorst}, \citenamefont {Deepu}, \citenamefont
  {Deger}, \citenamefont {Hahne}, \citenamefont {Sinha}, \citenamefont
  {Antonietti}, \citenamefont {Schmidt}, \citenamefont {Paz}, \citenamefont
  {Garrido}, \citenamefont {Ippen}, \citenamefont {Riquelme}, \citenamefont
  {Serenko}, \citenamefont {Kühn}, \citenamefont {Kitayama}, \citenamefont
  {Mørk}, \citenamefont {Spreizer}, \citenamefont {Jordan}, \citenamefont
  {Krishnan}, \citenamefont {Senden}, \citenamefont {Hagen}, \citenamefont
  {Shusharin}, \citenamefont {Vennemo}, \citenamefont {Rodarie}, \citenamefont
  {Morrison}, \citenamefont {Graber}, \citenamefont {Schuecker}, \citenamefont
  {Diaz}, \citenamefont {Zajzon},\ and\ \citenamefont {Plesser}}]{nest216}%
  \BibitemOpen
  \bibfield  {author} {\bibinfo {author} {\bibfnamefont {C.}~\bibnamefont
  {Linssen}}, \bibinfo {author} {\bibfnamefont {M.~E.}\ \bibnamefont
  {Lepperød}}, \bibinfo {author} {\bibfnamefont {J.}~\bibnamefont {Mitchell}},
  \bibinfo {author} {\bibfnamefont {J.}~\bibnamefont {Pronold}}, \bibinfo
  {author} {\bibfnamefont {J.~M.}\ \bibnamefont {Eppler}}, \bibinfo {author}
  {\bibfnamefont {C.}~\bibnamefont {Keup}}, \bibinfo {author} {\bibfnamefont
  {A.}~\bibnamefont {Peyser}}, \bibinfo {author} {\bibfnamefont
  {S.}~\bibnamefont {Kunkel}}, \bibinfo {author} {\bibfnamefont
  {P.}~\bibnamefont {Weidel}}, \bibinfo {author} {\bibfnamefont
  {Y.}~\bibnamefont {Nodem}}, \bibinfo {author} {\bibfnamefont
  {D.}~\bibnamefont {Terhorst}}, \bibinfo {author} {\bibfnamefont
  {R.}~\bibnamefont {Deepu}}, \bibinfo {author} {\bibfnamefont
  {M.}~\bibnamefont {Deger}}, \bibinfo {author} {\bibfnamefont
  {J.}~\bibnamefont {Hahne}}, \bibinfo {author} {\bibfnamefont
  {A.}~\bibnamefont {Sinha}}, \bibinfo {author} {\bibfnamefont
  {A.}~\bibnamefont {Antonietti}}, \bibinfo {author} {\bibfnamefont
  {M.}~\bibnamefont {Schmidt}}, \bibinfo {author} {\bibfnamefont
  {L.}~\bibnamefont {Paz}}, \bibinfo {author} {\bibfnamefont {J.}~\bibnamefont
  {Garrido}}, \bibinfo {author} {\bibfnamefont {T.}~\bibnamefont {Ippen}},
  \bibinfo {author} {\bibfnamefont {L.}~\bibnamefont {Riquelme}}, \bibinfo
  {author} {\bibfnamefont {A.}~\bibnamefont {Serenko}}, \bibinfo {author}
  {\bibfnamefont {T.}~\bibnamefont {Kühn}}, \bibinfo {author} {\bibfnamefont
  {I.}~\bibnamefont {Kitayama}}, \bibinfo {author} {\bibfnamefont
  {H.}~\bibnamefont {Mørk}}, \bibinfo {author} {\bibfnamefont
  {S.}~\bibnamefont {Spreizer}}, \bibinfo {author} {\bibfnamefont
  {J.}~\bibnamefont {Jordan}}, \bibinfo {author} {\bibfnamefont
  {J.}~\bibnamefont {Krishnan}}, \bibinfo {author} {\bibfnamefont
  {M.}~\bibnamefont {Senden}}, \bibinfo {author} {\bibfnamefont
  {E.}~\bibnamefont {Hagen}}, \bibinfo {author} {\bibfnamefont
  {A.}~\bibnamefont {Shusharin}}, \bibinfo {author} {\bibfnamefont {S.~B.}\
  \bibnamefont {Vennemo}}, \bibinfo {author} {\bibfnamefont {D.}~\bibnamefont
  {Rodarie}}, \bibinfo {author} {\bibfnamefont {A.}~\bibnamefont {Morrison}},
  \bibinfo {author} {\bibfnamefont {S.}~\bibnamefont {Graber}}, \bibinfo
  {author} {\bibfnamefont {J.}~\bibnamefont {Schuecker}}, \bibinfo {author}
  {\bibfnamefont {S.}~\bibnamefont {Diaz}}, \bibinfo {author} {\bibfnamefont
  {B.}~\bibnamefont {Zajzon}}, \ and\ \bibinfo {author} {\bibfnamefont {H.~E.}\
  \bibnamefont {Plesser}},\ }\href {\doibase 10.5281/zenodo.1400175} {\enquote
  {\bibinfo {title} {Nest 2.16.0},}\ } (\bibinfo {year} {2018})\BibitemShut
  {NoStop}%
\bibitem [{\citenamefont {Vreeswijk}\ and\ \citenamefont
  {Sompolinsky}(1998)}]{vreeswijk1998chaotic}%
  \BibitemOpen
  \bibfield  {author} {\bibinfo {author} {\bibfnamefont {C.~v.}\ \bibnamefont
  {Vreeswijk}}\ and\ \bibinfo {author} {\bibfnamefont {H.}~\bibnamefont
  {Sompolinsky}},\ }\href@noop {} {\bibfield  {journal} {\bibinfo  {journal}
  {Neural computation}\ }\textbf {\bibinfo {volume} {10}},\ \bibinfo {pages}
  {1321} (\bibinfo {year} {1998})}\BibitemShut {NoStop}%
\bibitem [{\citenamefont {Luque}\ and\ \citenamefont
  {Sol{\'e}}(2000)}]{luque2000lyapunov}%
  \BibitemOpen
  \bibfield  {author} {\bibinfo {author} {\bibfnamefont {B.}~\bibnamefont
  {Luque}}\ and\ \bibinfo {author} {\bibfnamefont {R.~V.}\ \bibnamefont
  {Sol{\'e}}},\ }\href@noop {} {\bibfield  {journal} {\bibinfo  {journal}
  {Physica A: Statistical Mechanics and its Applications}\ }\textbf {\bibinfo
  {volume} {284}},\ \bibinfo {pages} {33} (\bibinfo {year} {2000})}\BibitemShut
  {NoStop}%
\end{thebibliography}%
%\bibliographystyle{phjcp}

%%%%%%%%%%%%%%%%%%%%%%%%%%%%%%%%%%%%%%%%%%%%%%%%%%%%%%%%%%%%%%%%%%%%%


\section*{Supplementary material}

# Supplementary Material: Edge of chaos and avalanches in neural networks with heavy-tailed synaptic weight distribution

Łukasz Kuśmierz,[1, *] Shun Ogawa,[1] and Taro Toyoizumi[1, 2]

[1]*Laboratory for Neural Computation and Adaptation, RIKEN Center for Brain Science, 2-1 Hirosawa, Wako, Saitama 351-0198, Japan*
[2]*Department of Mathematical Informatics, Graduate School of Information Science and Technology, The University of Tokyo, Tokyo 113-8656, Japan*

## CONTENTS

## I. BINARY GAUSSIAN NETWORK

### A. Fully connected network

For a general activation function it is convenient to describe the behavior of the Gaussian networks in terms of another order parameter,

$$q_0(t) = \frac{1}{N} \sum_i \phi(x_i(t))^2. \tag{S.1}$$

However, in the binary case analyzed here it is equivalent to $m(t)$. The corresponding dynamical mean-field equation [1] reads

$$m(t+1) = \frac{1}{2}\left[1 - \operatorname{erf}\left(\frac{\theta}{\sqrt{2m(t)}g}\right)\right], \tag{S.2}$$

where the error function is given by $\operatorname{erf}(x) = 2\pi^{-1/2} \int_0^x \exp(-z^2)dz$. Expanding the RHS of (S.2) around $m(t) \approx 0$ gives

$$m(t+1) = \sqrt{\frac{m(t)g^2}{2\pi\theta^2}} \exp\left(-\frac{\theta^2}{2m(t)g^2}\right)\left(1 + O\left(\frac{m(t)g^2}{\theta^2}\right)\right). \tag{S.3}$$

Due to the exponential factor $m(t+1) < m(t)$ for small enough $m(t)$, which proves that the fixed point $m(t) = 0$ is locally stable under the evolution (S.2). Since the quiescent state is always stable, the transition to chaos, if present, must be discontinuous. This is confirmed by the graphical inspection of (S.2) (Fig. S.1) and in computer simulations (Fig. S.2).

---

* nalewkoz@gmail.com

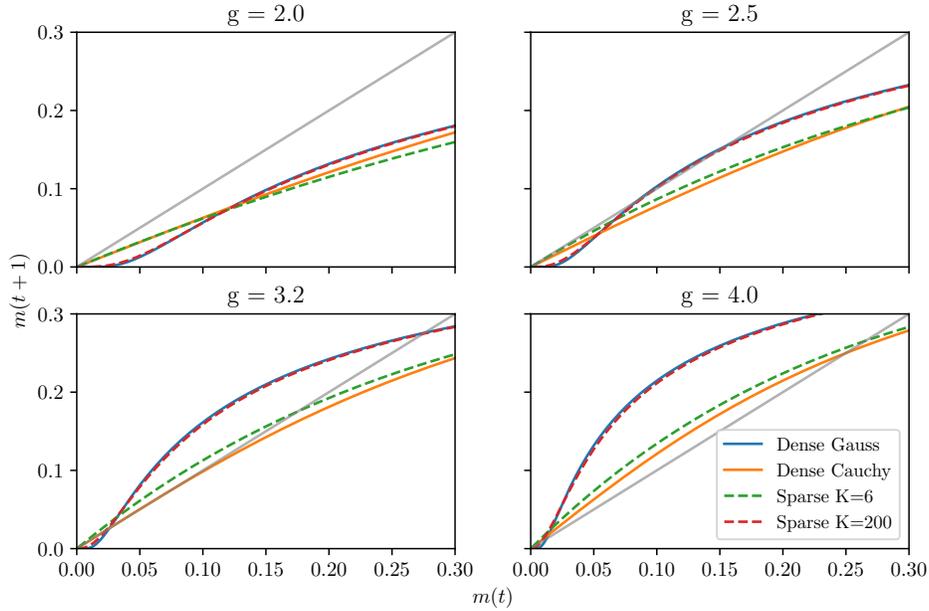

FIG. S.1. Mean field $m(t) \mapsto m(t+1)$ mapping for fully connected (dense) Gaussian and Cauchy networks and sparse Gaussian network with a fixed number of incoming connections per neuron $K$. For large values of $K$, the MF equation looks similar to the fully connected Gaussian case. In particular, we observe a discontinuous transition at $g \approx 2.5$. (note that this similarity only holds if $g$ is not too large, since for any finite $K$ there is a continuous transition at which the trivial fixed point loses its stability). For an intermediate sparsity ($2 < K \lessapprox 12$), only a second order transition is observed and the dynamics looks qualitatively similar to the dense Cauchy case. For $K \leq 2$ no transition to chaos is observed and the trivial fixed point is always stable (results not shown).

The mapping to a branching process offers a simple way of understanding this result. The probability that a given synapse is an autocrat reads

$$\text{Prob}(J_{ij} > \theta) = \frac{1}{2}\left(1 - \text{erf}\left(-\sqrt{\frac{N}{2}}\frac{\theta}{g}\right)\right) \tag{S.4}$$

In the thermodynamic limit the average number of autocrat connections per neuron can be calculated as before as $\lambda = \lim_{N \to \infty} NP(J_{ij} > \theta)$, which in our case leads to

$$\lambda = \lim_{N \to \infty} \sqrt{\frac{N}{2}}\frac{g}{\theta} \exp\left(-\frac{N\theta^2}{2g^2}\right)\left(1 + O(1/N^2)\right) = 0, \tag{S.5}$$

i.e. an activity starting from a single seed almost surely dies out.

Interestingly, the discontinuous character of the transition between quiescent and active phases have already been reported in various models that employ Gaussian weights [2, 3], which confirms the generic nature of this result.

### B. Sparse network

Let each neuron receive exactly $K$ incoming connections, randomly chosen from the network, and let $J_{ij} \sim \mathcal{N}(0, g^2/K)$. If the mean activity of the network at time $t$ is $m(t)$, the probability that exactly $n$ incoming neurons are active is given by the binomial distribution

$$P(n; m(t)) = \binom{K}{n} m(t)^n (1 - m(t))^{K-n}. \tag{S.6}$$

The membrane potential of a neuron $x$, conditioned on $n$ incoming connections being active, is a normal random variable with $\mu = 0$ and $\sigma^2 = ng^2/K$. The mean activity of the network in the next step is equal to the probability that the membrane potential of any given neuron crosses the threshold, and so it reads

$$m(t+1) = 2^{-1} \sum_{n=1}^{K} \binom{K}{n} m(t)^n (1-m(t))^{K-n} \left(1 - \text{erf}\left(\frac{\theta \sqrt{K}}{\sqrt{2n}g}\right)\right). \tag{S.7}$$

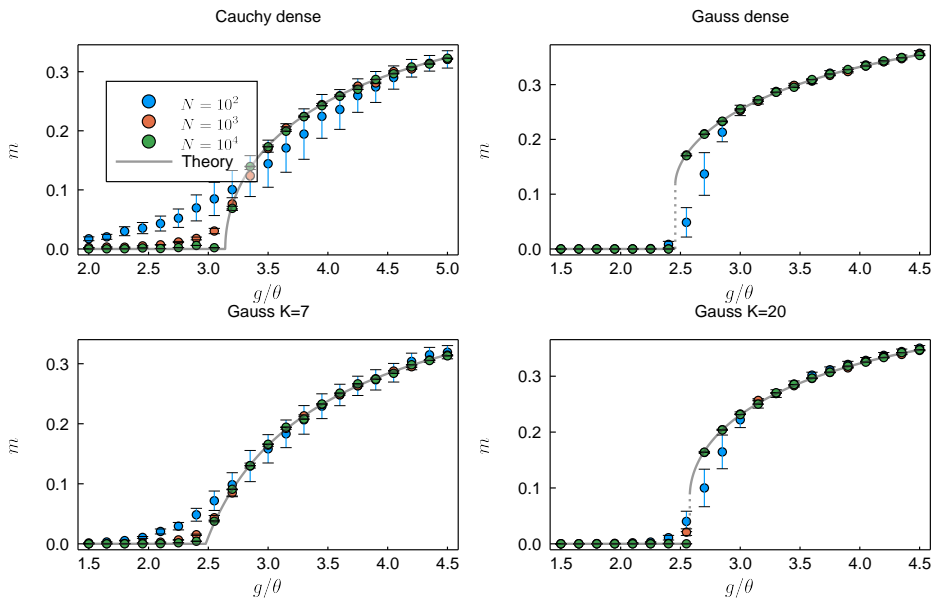

FIG. S.2. Steady-state mean activity as a function of the control parameter $g/\theta$. Lines were obtained by solving our mean field equations (8), (S.2), and (S.7) self-consistently. The points were obtained from computer simulations of $M = 10^5/N$ independent realizations of $\mathbf{J}$. For each realization of the weight matrix, the network was first evolved for 400 steps from an active state ($m(0) \approx 0.5$). The following 200 steps were used in calculating the average, which was performed over the steps, neurons, and realizations of $\mathbf{J}$. Error bars denote $\pm 3\sigma$ confidence intervals.

Our computer simulations corroborate the validity of (S.7), see Fig. S.2. Note, additionally, that it is not difficult to show that (S.7) simplifies to the dense case (S.2) in the limit of $K \to \infty$. Analogous calculations in the case of stochastic units were presented in [4], where the range of validity of the annealed approximation is also discussed.

## II. MEAN FIELD EQUATIONS CORRECTLY PREDICT THE MEAN ACTIVITY IN THE THERMODYNAMIC LIMIT

We have performed computer simulations of Cauchy, fully connected Gaussian, and sparse Gaussian networks with binary activation functions (Fig. S.2). The good match between our theoretical predictions and the simulation results strongly suggests that the mean field approach is exact in the thermodynamic limit.

## III. ACTIVE PHASE IS CHAOTIC IN RANDOM NETWORKS OF BINARY NEURONS

Here we show, with a simple mathematical argument and computer simulations, that whenever the mean network activity $m$ is non-zero in the steady-state, the corresponding attractor is chaotic. This is true for both Gaussian and Cauchy networks with the binary activation function, and is expected to hold in more general settings, unless some non-trivial structure of the connectivity is present, e.g. symmetric or antisymmetric connectivity (the latter with a symmetric version of the binary activation function, i.e. with $\theta = 0$ and taking values $\pm 1$) leads to non-chaotic, periodic attractors [5].

It is easy to prove our result in the Gaussian case. In this case, with $N \to \infty$, the maximal Lyapunov exponent is given by [1, 6, 7]

$$\lambda_{\max}^{\text{gauss}} = \frac{1}{2} \ln g^2 \int_{-\infty}^{\infty} Dx \left[\phi'\left(g\sqrt{q_0}x\right)\right]^2 \tag{S.8}$$

where $Dx$ denotes the standard Gaussian measure, $q_0$ is the average squared activity (S.1). It is clear that whenever $\phi$ features any discontinuity and the phase is active ($g\sqrt{q_0} \neq 0$), formula (S.8) predicts $\lambda_{\max}^{\text{gauss}} = \infty$. In particular, the binary activation function has a discontinuity at $x = \theta$ and thus the corresponding active phase is always chaotic. The analysis for the binary Cauchy network is more involved and will be published elsewhere. However, the intuition is the same: small perturbations in the active phase are expanded due to the discontinuity of the activation function. Note that these results are in line with previous

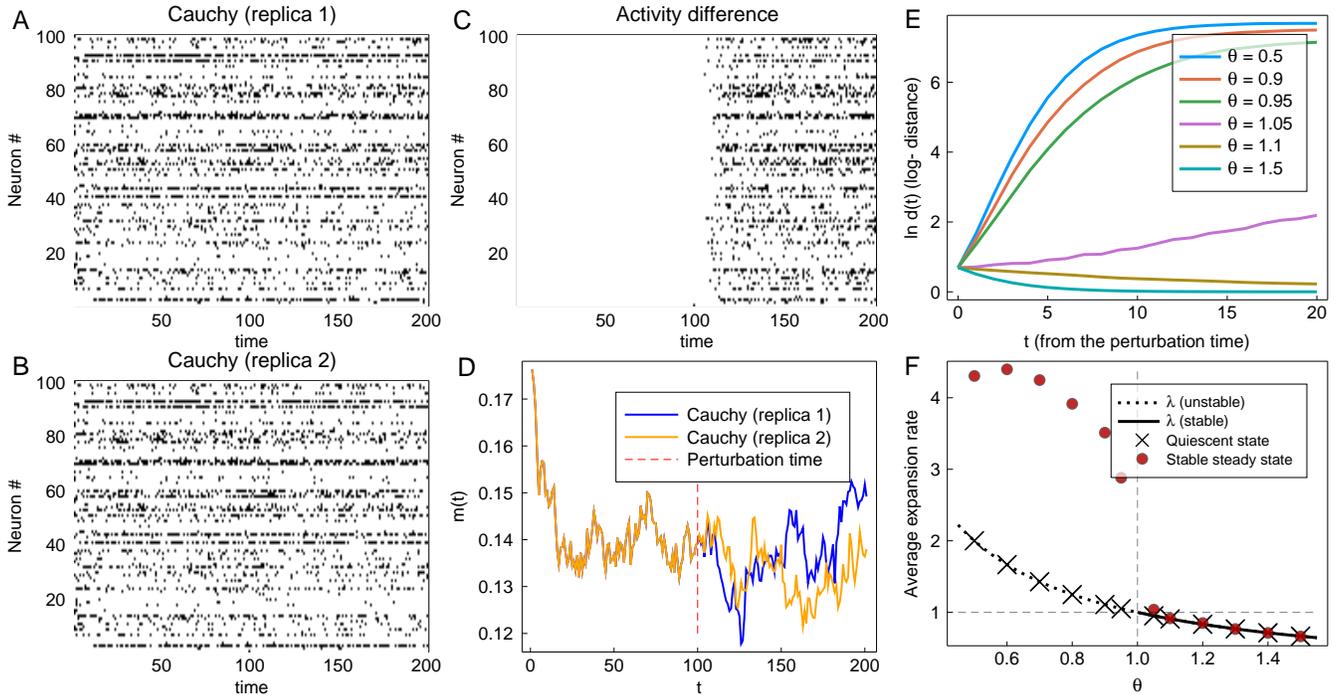

FIG. S.3. Numerical perturbation analysis of the binary Cauchy network. (A-D): Two replicas were evolved from identical initial conditions ($\theta = 0.95$). A single-neuron perturbation, introduced at $T_0 = 100$, is quickly expanded, which can be seen by comparing individual neuron activities (C) or mean network activities (D). (E): The average Hamming distances between perturbed and unperturbed states as functions of the number of steps $T$ after introducing a single-neuron state flip to a steady-state trajectory. The network is sensitive to the perturbations for $\theta < 1$, corroborating the existence of a transition to chaos at the predicted value of $\theta = 1$. (F): The average expansion rate as a function of $\theta$. As expected, the average expansion rate from the quiescent initial state is equal to $\lambda = g/(\pi\theta)$. The average expansion rate from the steady state is different from $\lambda$ for $\theta < 1$; in this regime the quiescent state does not correspond to the stable steady state (the quiescent state is unstable). In (E) and (F) the distances are averaged over all possible neurons and over 10 realizations of $\boldsymbol{J}$. See text for more details.

studies, e.g., it has been shown that in random Boolean networks Lyapunov exponents scale as $\ln K$ with the number of incoming connections [8]. The issue of infinite Lyapunov exponents in binary networks is also discussed in detail in [9].

To support our claim we simulated Cauchy networks with binary activation functions. Two replicas that shared $\boldsymbol{J}$ were evolved from identical initial conditions. At a fixed time $T_0 = 100$ a perturbation was introduced by flipping the state of a single neuron in one of the replicas. We then calculate the activity difference

$$\delta_i(t) = \left|\phi\left(x_i^1(t)\right) - \phi\left(x_i^2(t)\right)\right|, \tag{S.9}$$

the Hamming distance

$$d(t) = \sum_{i=1}^{N} \delta_i(t), \tag{S.10}$$

and the average expansion rate

$$r = \frac{d(T_0 + 1)}{d(T_0)}, \tag{S.11}$$

where $d(T_0) = 1$. Similarly, we calculated the average expansion rate from the quiescent initial state $\lambda$. In this case we evolved the (unperturbed) network from a quiescent initial state, which corresponds to the stable steady state for $\theta > \theta^* = 1$ ($g = \pi$), whereas for $\theta < 1$ is unstable. The theoretically calculated branching parameter agrees with $\lambda$ obtained from the simulations (Fig. S.3F). The evolution of the network is sensitive to the perturbation for $\theta < 1$, which can be observed at the level of individual neuron's activities (Fig. S.3A-C) and, due to the finite size of the network, the mean activity (Fig. S.3D). In the chaotic phase, the Hamming distance initially grows exponentially and then saturates at $d(\infty) \sim N$. The chaotic phase is signaled by $r > 1$, which is directly related to the maximal Lyapunov exponent being positive [8–10]. Although $r$ in general is not equal to $\lambda$, both quantities cross the value 1 at the critical point $\theta = 1$. The discrepancy observed around the transition point (the slowly growing distance at $\theta = 1.05$ in Fig. S.3E and the corresponding $r \approx 1.04 > 1$ (Fig. S.3F)) is expected to be a finite size effect.

## IV. DETAILS OF THE SIMULATIONS

Computer simulations of binary networks were performed using custom-written codes in MATLAB and Julia [11]. The results for spiking neural networks were obtained with NEST Simulator [12] through a custom-written code in Python. Details of the simulations:

**Fig. 2**: Networks had size $N = 10^4$ and results were averaged over 10 independent realizations of $\mathbf{J}$. For a given realization of $\mathbf{J}$, we run $10^4$ simulations, each starting from different seed neuron (i.e. one neuron active, all other neurons inactive). $g = \pi$.

**Fig. 3**: Both Gaussian and Cauchy networks were fully connected with $N = 10^4$. The injected current changed between −400 pA and 400 pA in small increments every 5 ms. All neurons were of type *iaf_psc_alpha*, which denotes a leaky integrate-and-fire neuron with alpha-shaped postsynaptic currents. Default parameters of the model neuron were used, i.e.: resting potential $E_L = -70$ mV, capacity of the membrane $C_m = 250$ pF, membrane time constant $\tau_m = 10$ ms, refractory period $t_{ref} = 2$ ms, spike threshold $V_{th} = -55$ mV, reset potential $V_{reset} = -70$ mV, rise time of the excitatory and inhibitory synaptic alpha function $\tau_{syn} = 2$ ms. Static synapses were used with the default delay of 1 ms, and weights were randomly drawn from symmetric (a) Gaussian distribution with $\sigma = 2.4 \times 10^3 / \sqrt{N}$ pA, and (b) Cauchy distribution with $\gamma = 1.92 \times 10^3/N$ pA. Poisson noise was injected randomly into the network, activating each neuron approximately twice every second. Note that without any external input the analyzed networks cannot recover from the quiescent state, since the model neurons are never spontaneously active. Sub-sampling: For the sake of clarity and drawing efficiency, in the raster plots only 10% of spikes of 100 randomly chosen neurons were drawn. Activity histograms were created using all data.

**Fig. S.3**: The averaging was performed over all $N = 10^4$ neurons (i.e., including the unperturbed network, $N + 1$ replicas were simulated) and over 10 realizations of $\mathbf{J}$. At time $T = 0$ the unperturbed network was prepared in the (a) steady state by evolving it for 100 steps before introducing perturbations or (b) quiescent state. $g = \pi$. Sub-sampling: For the sake of clarity and drawing efficiency, activities of a randomly chosen subpopulation of neurons ($N' = 100$) is shown.

## V. AVALANCHE STATISTICS

The mapping to the branching process together with the general results known for critical branching processes provide the critical exponents [13–15]. Here, for completeness, we calculate two critical exponents in our specific case.

### A. Avalanche size distribution

The size of an avalanche is defined as the sum of the number of active neurons at each time step, from the beginning of the avalanche till its end. Let $S_m$ denote the size of an avalanche starting from $m$ seeds and $G_m(z)$ denote the corresponding generating function

$$G_m(z) \equiv \langle z^{S_m} \rangle = \sum_{j=0}^{\infty} z^j \, \text{Prob}(S_m = j). \tag{S.12}$$

Since the activity of network is assumed to be sparse, the avalanche that starts from $m$ seeds consists of $m$ independent avalanches starting from a single seed. Therefore we can write that $S_m = \sum_{i=1}^{m} S_1^{(i)}$, where $\{S_1^{(i)}\}$ is a set of i.i.d. random variables denoting sizes of single-seed generated avalanches. At the level of the generating functions, this assumption leads to a simple expression

$$G_m(z) = [G_1(z)]^m. \tag{S.13}$$

On the other hand, we know how a seed neuron propagates the activity through the network in a single step: it activates $m$ neurons with probability

$$p_m = \frac{\lambda^m}{m!} e^{-\lambda}, \tag{S.14}$$

where $\lambda$ is given by (10). This means that one seed generate an avalanche of size $1 + S_m$ with probability $p_m$, where 1 is from the first step (i.e., the seed) and $S_m$ is from the subsequent steps. Hence, the single-seed generating function can be calculated as follows

$$G_1(z) = \sum_{m=0}^{\infty} p_m \langle z^{1+S_m} \rangle = z \sum_{m=0}^{\infty} p_m G_m(z). \tag{S.15}$$

We combine (S.15) with (S.13) and (S.14) and arrive at an implicit expression for the one-seed avalanche size generating function

$$G_1(z) = z\left\{p_0 e^{S_0} + \sum_{m=1}^{\infty} p_m [G_1(z)]^m\right\} = z\exp(\lambda G_1(z) - \lambda). \tag{S.16}$$

Note that we used $S_0 = 0$ (no avalanche without a seed). In order to inspect the tail of the distribution of $S_1$ we introduce an auxiliary function $g(\epsilon) = 1 - G_1(1 - \epsilon)$ and expand the RHS of (S.16) assuming that $\epsilon \ll 1$ and $g(\epsilon) \ll 1$ (valid for $\lambda \leq 1$):

$$1 - g(\epsilon) = (1 - \epsilon)\left(1 - \lambda g(\epsilon) + \frac{\lambda^2 g(\epsilon)^2}{2} + O(g(\epsilon)^3)\right), \tag{S.17}$$

which to the lowest order can be rewritten as

$$g(\epsilon) = \begin{cases} \frac{\epsilon}{1-\lambda}, & \text{for } \lambda \neq 1 \\ \sqrt{2\epsilon}, & \text{for } \lambda = 1. \end{cases} \tag{S.18}$$

The small $\epsilon$ behavior of $g(\epsilon) \sim \epsilon^{1/2}$ in the $\lambda = 1$ case translates into the tail behavior of the avalanche size density as

$$\text{Prob}(S_1 = s) \sim s^{-3/2} \tag{S.19}$$

for large $s$.

### B. Avalanche lifetime distribution

Let $T_m$ be the lifetime (number of steps with nonzero activity) of an avalanche that starts from $m$ seeds. By definition $T_0 = 0$ and $T_m \geq 1$ for $m > 0$. As before, we treat an avalanche from different seeds as independent, and thus the following identity linking the survival probabilities holds

$$Q_m(t) = 1 - \text{Prob}(T_m \leq t) = 1 - [\text{Prob}(T_1 \leq t)]^m = 1 - [1 - Q_1(t)]^m. \tag{S.20}$$

As in the case of the size distribution, we can unwrap the first step of the dynamics starting from a single seed, which gives

$$Q_1(t+1) = \sum_{m=1}^{\infty} p_m Q_m(t) = 1 - \sum_{m=0}^{\infty} p_m [1 - Q_m(t)]. \tag{S.21}$$

We plug (S.20) and (S.14) into (S.21) and arrive at the following recursive relation

$$Q_1(t+1) = 1 - \exp[-\lambda Q_1(t)]. \tag{S.22}$$

If $\lambda > 1$ there exists a non-zero fixed point corresponding to the non-zero probability of survival at $t \to \infty$. In contrast, the activity eventually dies out almost surely for $\lambda \leq 1$. Assuming $\lambda Q_1(t) \ll 1$ the recursive relation simplifies to

$$Q_1(t+1) = \lambda Q_1(t) - \frac{\lambda^2}{2} Q_1(t)^2, \tag{S.23}$$

which predicts an exponential decay for $\lambda < 1$. At the critical point $\lambda = 1$ and the decay is a power law. In that case the recursion can be solved with an ansatz $Q_1(t) = C/t^{\delta}$, leading to $\delta = 1$, as expected.

## VI. EXISTENCE OF A CONTINUOUS TRANSITION IN CAUCHY NETWORKS WITH A POSITIVE THRESHOLD

Let $\phi(x)$ be an activation function such that $\phi(x) = 0$ for $x$ below a positive threshold $\theta$, and $\phi(x) \approx C$ for sufficiently large $x$ (i.e. $x > m_1$). Without much loss of generality, we additionally assume that $\phi(x) \geq 0$ and $\int_a^b \phi(x)dx < \infty$. The integral in the mean-field equation can be then decomposed as follows:

$$m(t+1) = \frac{1}{\pi}\int_0^{\infty} dz \frac{\phi(m(t)gz)}{1+z^2} \approx \frac{1}{\pi}\int_{\frac{\theta}{m(t)g}}^{\frac{m_1}{m(t)g}} dz \frac{\phi(m(t)gz)}{1+z^2} + \frac{C}{\pi}\int_{\frac{m_1}{m(t)g}}^{\infty} dz \frac{1}{1+z^2} = \frac{m(t)g}{\pi}\int_{\theta}^{m_1} dy \frac{\phi(y)}{[m(t)g]^2 + y^2} + \frac{C}{\pi}\arctan\left(\frac{m(t)g}{m_1}\right). \tag{S.24}$$

We expand (S.24) around $m(t) = 0$:

$$m(t+1) = \frac{m(t)g}{\pi} \left( \int_\theta^{m_1} dy \frac{\phi(y)}{y^2} + \frac{C}{m_1} \right) + O(m(t)^3), \qquad (S.25)$$

and conclude that the transition between quiescent and active state occurs at the critical point described by the equation

$$\frac{g}{\pi} \left( \int_\theta^{m_1} dy \frac{\phi(y)}{y^2} + \frac{C}{m_1} \right) = 1. \qquad (S.26)$$

Hence, this guarantees the existence of positive and finite critical $g^*$ that solves the above equation. It is possible to extend these results to activation functions $\phi(x)$ that are non-zero around $x = 0$ and are non-saturating: the transition exists if $\phi(x)$ is sufficiently superlinear around $x = 0$ and sublinear for large $|x|$.

---

[1] L. Molgedey, J. Schuchhardt, and H. G. Schuster, Physical review letters **69**, 3717 (1992).
[2] N. Brunel, Journal of computational neuroscience **8**, 183 (2000).
[3] J. H. Marshel, Y. S. Kim, T. A. Machado, S. Quirin, B. Benson, J. Kadmon, C. Raja, A. Chibukhchyan, C. Ramakrishnan, M. Inoue, *et al.*, Science **365**, eaaw5202 (2019).
[4] B. Derrida, E. Gardner, and A. Zippelius, EPL (Europhysics Letters) **4**, 167 (1987).
[5] E. Goles, Discrete Applied Mathematics **13**, 97 (1986).
[6] M. Massar and S. Massar, Physical Review E **87**, 042809 (2013).
[7] G. Wainrib and M. N. Galtier, Neural Networks **76**, 39 (2016).
[8] B. Luque and R. V. Solé, Physica A: Statistical Mechanics and its Applications **284**, 33 (2000).
[9] C. v. Vreeswijk and H. Sompolinsky, Neural computation **10**, 1321 (1998).
[10] L. Büsing, B. Schrauwen, and R. Legenstein, Neural computation **22**, 1272 (2010).
[11] J. Bezanson, A. Edelman, S. Karpinski, and V. B. Shah, SIAM review **59**, 65 (2017).
[12] C. Linssen, M. E. Lepperød, J. Mitchell, J. Pronold, J. M. Eppler, C. Keup, A. Peyser, S. Kunkel, P. Weidel, Y. Nodem, D. Terhorst, R. Deepu, M. Deger, J. Hahne, A. Sinha, A. Antonietti, M. Schmidt, L. Paz, J. Garrido, T. Ippen, L. Riquelme, A. Serenko, T. Kühn, I. Kitayama, H. Mørk, S. Spreizer, J. Jordan, J. Krishnan, M. Senden, E. Hagen, A. Shusharin, S. B. Vennemo, D. Rodarie, A. Morrison, S. Graber, J. Schuecker, S. Diaz, B. Zajzon, and H. E. Plesser, "Nest 2.16.0," (2018).
[13] P. Alstrøm, Physical Review A **38**, 4905 (1988).
[14] M. A. Munoz, R. Dickman, A. Vespignani, and S. Zapperi, Physical Review E **59**, 6175 (1999).
[15] G. Ódor, Reviews of modern physics **76**, 663 (2004).



\end{document}